\documentclass[journal,twoside]{IEEEtran}
\IEEEoverridecommandlockouts
\usepackage{amsmath,amssymb,amsfonts}
\usepackage{algorithmic}
\usepackage{graphicx}
\usepackage{textcomp}
\usepackage{xcolor}
\usepackage{gensymb}
\usepackage{multirow, multicol}
\usepackage[font=footnotesize]{caption}
\usepackage[font=footnotesize]{subcaption}
\usepackage{graphicx} 
\usepackage{makecell}
\usepackage[detect-all]{siunitx}
\usepackage[acronym]{glossaries-prefix}
\usepackage{cite}

\usepackage{array}
\usepackage{booktabs}
\usepackage{flushend}
\newcommand{\mycomment}[1]{}
\usepackage{bm}
\usepackage{etoolbox}
\usepackage{diagbox}
\usepackage{newtxmath}
\makeatletter

\newacronym[plural={LWAs},prefixfirst={a\ }, prefix={an\ }]{lwa}{LWA}{leaky-wave antenna}
\newacronym{osb}{OSB}{open stopband}
\newacronym{lh}{LH}{left-handed}
\newacronym{crlh}{CRLH}{composite right left-handed}
\newacronym{siw}{SIW}{substrate integrated waveguide}
\newacronym{tem}{TEM}{transverse electromagnetic}
\newacronym{vna}{VNA}{vector network analyzer}
\newacronym{fmcw}{FMCW}{frequency-modulated continuous wave}
\newacronym{fov}{FoV}{field of view}
\newacronym{em}{EM}{electromagnetic}
\newacronym{rf}{RF}{radio frequency}
\newacronym{mmw}{mmWave}{millimeter-wave}
\newacronym{DoA}{DoA}{Direction of Arrival}
\newacronym{sws}{SWS}{slow wave structure}
\newacronym{pcb}{PCB}{printed circuit board}
\newacronym{hpbw}{HPBW}{half power beam-width}
\newacronym{catr}{CATR}{compact antenna test range}
\newacronym{xpd}{XPD}{cross-polarization discrimination}
\newacronym{bw}{BW}{bandwidth}
\newacronym{sspp}{SSPP}{spoof surface plasmon polariton}
\newacronym{ebg}{EBG}{electromagnetic bandgap}
\newacronym{apd}{APD}{absorbed power density}
\newacronym{sar}{SAR}{specific absorption rate}
\newacronym{sc}{SC}{stratum corneum}
\newacronym{br}{BR}{basic restriction}
\newacronym{srd}{SRD}{short range devices}
\newacronym{tap}{TAP}{total absorbed power}
\newacronym{ban}{BAN}{body area network}
\newacronym{eirp}{EIRP}{equivalent isotropic radiated power}

\date{ }

\def\BibTeX{{\rm B\kern-.05em{\sc i\kern-.025em b}\kern-.08em
    T\kern-.1667em\lower.7ex\hbox{E}\kern-.125emX}}
\begin{document}

\title{Conformal Wide-Angle Scanning Leaky-Wave Antenna for V-Band On-Body Applications}

\author{Pratik~Vadher,~\IEEEmembership{Student Member,~IEEE,}
Anja K. Skrivervik,
Qihang~Zeng,~\IEEEmembership{Member,~IEEE,}
Ronan~Sauleau,~\IEEEmembership{Fellow,~IEEE,}
John S. Ho,~\IEEEmembership{Member,~IEEE,}
Giulia~Sacco,~\IEEEmembership{Member,~IEEE,}
and~Denys~Nikolayev,~\IEEEmembership{Senior Member,~IEEE}%

\thanks{Manuscript received July XX, 2024, revised XX X, 2024.}%
\thanks{This project has received funding from the French Agence Nationale de la Recherche (ANR) through the Project MedWave under Grant ANR-21-CE19-0045; from the European Union's Horizon Europe research and innovation program through the Marie Sk\l odowska-Curie IN-SIGHT project N$^\circ$101063966, European Research Council BESSEL project N$^\circ$101165786, and from the French National Research Agency (ANR-22-PEFT-0007) as part of France 2030 and the NF-FITNESS project. \textit{(Corresponding authors: Denys Nikolayev, denys.nikolayev@deniq.com; Giulia Sacco, giulia.sacco@cnrs.fr)}}%
\thanks{This work is supported by the European Union through European Regional Development Fund (ERDF), Ministry of Higher Education and Research, CNRS, Brittany region, Conseils Départementaux d’Ille-et-Vilaine and Côtes d’Armor, Rennes Métropole, and Lannion Trégor Communauté, through the CPER Project CyMoCod.}
\thanks{P. Vadher, R. Sauleau, G. Sacco, and D. Nikolayev are with the IETR~-- UMR 6164, CNRS / Univ Rennes, FR-35000 Rennes, France.}%
\thanks{A. K. Skrivervik is with the Microwave and Antenna Group, Ecole Polytechnique Fédérale de Lausanne, CH-1015 Lausanne, Switzerland.}%
\thanks{Q. Zeng, and J. S. Ho are with the Department of Electrical and Computer Engineering, the Institute for Health Innovation and Technology, and the N.1 Institute, National University of Singapore, Singapore 117583}
}%


\maketitle

\begin{abstract}
Wearable on-body \gls{mmw} radars can provide obstacle detection and guidance for visually impaired individuals. \textcolor{black}{The antennas being a crucial component of these systems, must be lightweight, flexible, low-cost, and compact. However, existing antennas suffer from a rigid form factor and limited reconfigurability.} This article presents a low-profile, fast scanning \gls{lwa} operating in the unlicensed V-band (\SIrange[range-units=single, range-phrase=--]{57}{64}{\GHz}) for on-body applications such as lightweight portable \gls{fmcw} radars. \textcolor{black}{The novel meandering microstrip design allows independent control of gain and scanning rate (rate of change of main beam pointing direction with frequency)}. Experimental results show that the \gls{lwa} achieves a realized gain above \SI{10}{\dB} with a fan-beam steering range in the H-plane from \SI{-35}{\degree} to \SI{45} {\degree} over the operating frequency band, while the \gls{hpbw} is within \SI{20}{\degree} in planar condition. To assess the on-body applicability, the antenna’s performance is evaluated under bending. When placed on the knee (corresponding to \SI{80}{\milli\meter} radius), the beam steers from \SI{-25}{\degree} to \SI{55}{\degree} with a maximum realized gain degradation of \SI{1.75}{\dB}, and an increase of \gls{hpbw} up to \SI{25}{\degree}. This demonstrates the \gls{lwa}'s robustness in conformal conditions, while maintaining beam-forming and beam-scanning capabilities. \textcolor{black}{Simulations confirm that the \gls{lwa}'s ground plane minimizes user exposure, adhering to international guidelines.} Finally, we demonstrate a 2-D spatial scanning by employing an array of twelve \glspl{lwa} with phased excitation, enabling beam-forming in the E-plane from \SI{-50}{\degree} to \SI{50} {\degree}, while the \gls{hpbw} remains below \SI{20}{\degree}. \textcolor{black}{Mutual coupling analysis reveals that isolation loss and active reflection coefficient remain below \SI{15}{\dB} throughout the operating band.}
\end{abstract}
\glsresetall
\glsresetall

\section{Introduction}
Enhancing navigation aids for visually impaired people is essential to improve their mobility and independence. Recent advances in mobile technology have led to an increased interest in developing wearable devices as assistive tools for people with visual impairments \cite{cardillo_millimeter-wave_2022, kurata_indoor-outdoor_2011, mahmud_vibration_2014, islam_developing_2019}. Of the various sensing modalities,  radars offer a cost-effective, lightweight, and robust sensing solution, since they do not depend on ambient light conditions and do not affect the user's privacy. In particular, short-range \gls{fmcw} radars can be implemented in a compact electronic system due to their low power consumption and low profile \cite{shoykhetbrod_scanning_2014, cardillo_millimeter-wave_2022}. Traditionally, these sensors have been placed on rigid objects (e.g., white canes and engineered glasses) which can burden the user \cite{cardillo_millimeter-wave_2022, bhatlawande_design_2014}. An improved solution consists of placing the \textcolor{black}{sensor} on the user's body in a way that does not hamper their movements and that is comfortable to wear \cite{kwiatkowski_concept_2017}.

For this application, the unlicensed V-band extending from \SIrange{57}{64}{\GHz} is a promising solution, since it provides faster communication \cite{5936164}, high range resolution, while being less susceptible to interference from nearby devices \cite{cotton_simulated_2010}. At these frequencies, when positioning the antenna on the user's body, its compact size reduces the impact of body curvature (radius ranging from \SIrange{40}{240}{\milli\meter} \cite{rahmat-samii_advances_2021}) by limiting bending-related distortions.

\begin{figure*}[!t]\centerline{\includegraphics[width=0.72\textwidth]{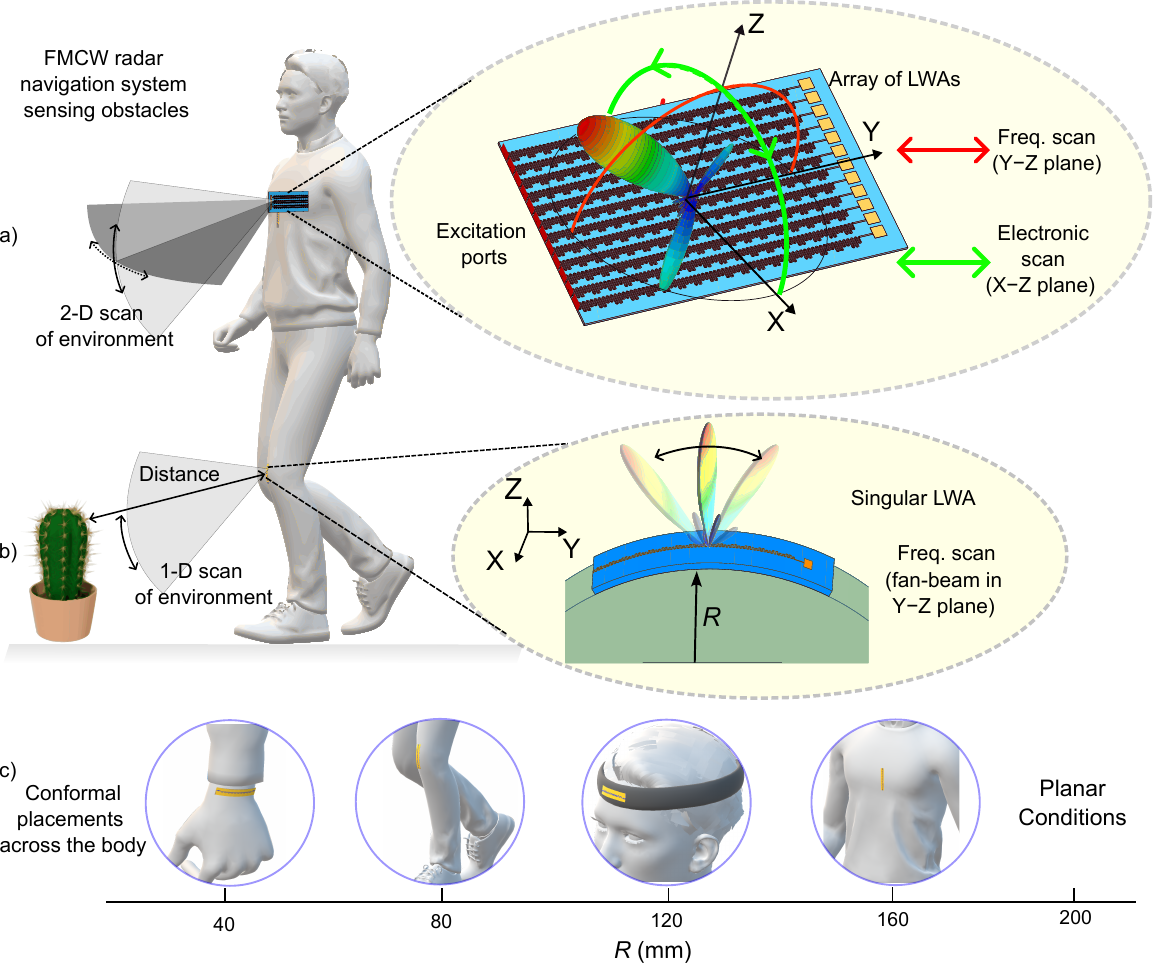}}
\caption{On-body FMCW V-band radar for sensing and detecting obstacles in the outdoor environment. (a)~Array of 12 LWAs. The scan in the Y--Z plane is due to frequency scanning, while the scanning in X--Z plane is due to electronic scanning. (b)~A single LWA and radiated fan-beam in the Y--Z plane. (c)~Possible antenna placements and corresponding curvature radii.}
\label{FigureOne}
\end{figure*}

\textcolor{black}{For conformal radar applications requiring localization, achieving both high angular resolution and broad angular coverage is essential \cite{nikolayev_conformal_2025, shoykhetbrod_scanning_2014, shoykhetbrod_design_2012}. This necessitates the use of pattern-reconfigurable antennas. While antenna arrays provide a straightforward solution, they demand additional hardware, including phase shifters and complex feeding networks \cite{azari2024design, eid20215g, su_compact_2024}. Moreover, at V-band frequencies, the inductive and capacitive losses associated with active and passive components---such as phase shifters, diodes, and varactors---significantly degrade performance, making them less attractive for practical implementation \cite{wang_metamaterial-based_2022, zandamela_3-d_2024} at V-band frequencies}. In contrast, \glspl{lwa} can be fed very easily and provide the steering of the main beam with frequency due to the dispersive nature of the guiding medium \cite{oliner_guided_1959, zheng_leaky-wave_2023, balanis_antenna_2015}. The use of \glspl{lwa} in combination with \gls{fmcw} radars represents a promising solution to limit the number of channels and the use of complex phase shifters \cite{sacco_siso_2023, mercuri_2-d_2021}. \Glspl{lwa} proposed earlier for such applications in the V-band \cite{vadher_-body_2022, shoykhetbrod_design_2012} have a large scanning range, but are characterized by a low rate of change of the main beam pointing direction with frequency, i.e., a low beam steering rate ($S_\textrm{m}$). To limit spectral resources for a given scanning range \cite{zheng_leaky-wave_2023,zhang_high-scanning-rate_2023,gomez-tornero_smart_2022}, a fast $S_\textrm{m}$ is required.

Furthermore, since the antenna is meant to be placed conformally on the user's body, it is of paramount importance that it is flexible ~\cite{soh_wearable_2015, song_systematic_2018}. To improve flexibility and reduce manufacturing costs, microstrip-based \glspl{lwa} \cite{vadher_meandering_2024, vadher_-body_2022, vadher_higher_2023, cheng_approximate_2019} are preferable to \gls{siw} based \glspl{lwa} \cite{sarkar_compact_2020, caloz_crlh_2008, liu_substrate_2012} given the absence of via holes. Additionally, periodic meandering microstrips \glspl{lwa} provide a 
good control over the path taken by the \gls{em} wave in the guiding medium, since the microstrip intervals between the corners---corresponding to the radiating discontinuities in this type of \gls{lwa}---can be tailored to ensure optimal polarization and scanning range of the main fan beam \cite{vadher_meandering_2024, wood_curved_1979, hall_microstrip_1983}.

In this paper, we propose and demonstrate a conformal \gls{lwa} that can be fabricated using a single printed layer and provides wide-scanning range from \SI{-35}{\degree} to \SI{45} {\degree} and \SI{-25}{\degree} to \SI{55}{\degree} when deformed on a curved surface over the operating range. The unit cell is composed of 4 meanders of varying vertical length. The microstrip-based design of the \gls{lwa} provides an effective control over the size of unit cell and hence over $S_\textrm{m}$. The length of the meandering microstrip intervals between the mitred corners (radiating discontinuities) are tailored to improve the \gls{xpd} ratio over a wide operation band \SIrange{57}{64}{\GHz}.
 
The paper is structured as follows. Section \ref{Operational_Principle_and_Antenna_Requirements} describes the proposed antenna system and the possible conformal wearable implementations. Section \ref{SectionMainTheory} explains the design process and the array-based model for the proposed \gls{lwa} starting from the modification of a \gls{sws} to obtain radiation, while eliminating the \gls{osb}. Section \ref{SectionPrototype} describes the design of the feeding structure and the measurements of the designed prototype in planar and conformal environments. Section \ref{sec: 2-D scanning} proposes a possible configuration of an array of 12 \glspl{lwa} to enable 2-D scanning. Section \ref{SectionConclusion} contains the comparison with state-of-the-art followed by conclusions.

\section{Operational Principle\\ and Antenna Requirements}\label{Operational_Principle_and_Antenna_Requirements}
\mycomment{We describe an example use case of the proposed on-body \gls{lwa}.} A lightweight wearable on-body radar should be able to detect and warn the users of approaching obstacles, as well as to create a map of their surroundings to aid their navigation [Fig.~\ref{FigureOne}~(a,b)]. The proposed antenna system is designed to be used in combination with an \gls{fmcw} radar positioned on the user's body surface. To enable 2-D scanning, an array of twelve \Glspl{lwa} is used as shown in Fig.~\ref{FigureOne}(a). The scanning in the Y--Z plane is ensured by the leaky-wave phenomenon, while the beamforming in the X--Z plane is obtained electronically, by modifying the excitation of the different \glspl{lwa} \cite{orth_novel_2019, kwiatkowski_combining_2024}. \textcolor{black}{Based on available commercial radar devices operating in the \SI{60}{\giga\hertz} band \cite{BGT60LTR11AIP}, a scanning range of $\SI{90}{\degree}$ can be considered promising for the proposed application.} To improve the wearability of the device, the antenna needs to be positioned conformally to the body surface and this is achieved thanks to the use of flexible substrates, as shown in Fig.~\ref{FigureOne}(b). Some possible locations along with the typical curvature radii \cite{rahmat-samii_advances_2021, wagih_millimeter-wave_2019} are shown in Fig.~\ref{FigureOne}(c). Finally, some precautions to minimize the exposure of the user such as the use of a ground plane are required \cite{vadher_-body_2022}. 

\section{Leaky-Wave theory for the Proposed Antenna}\label{SectionMainTheory}

\begin{figure}[!t]\centerline{\includegraphics[width=0.4\textwidth]{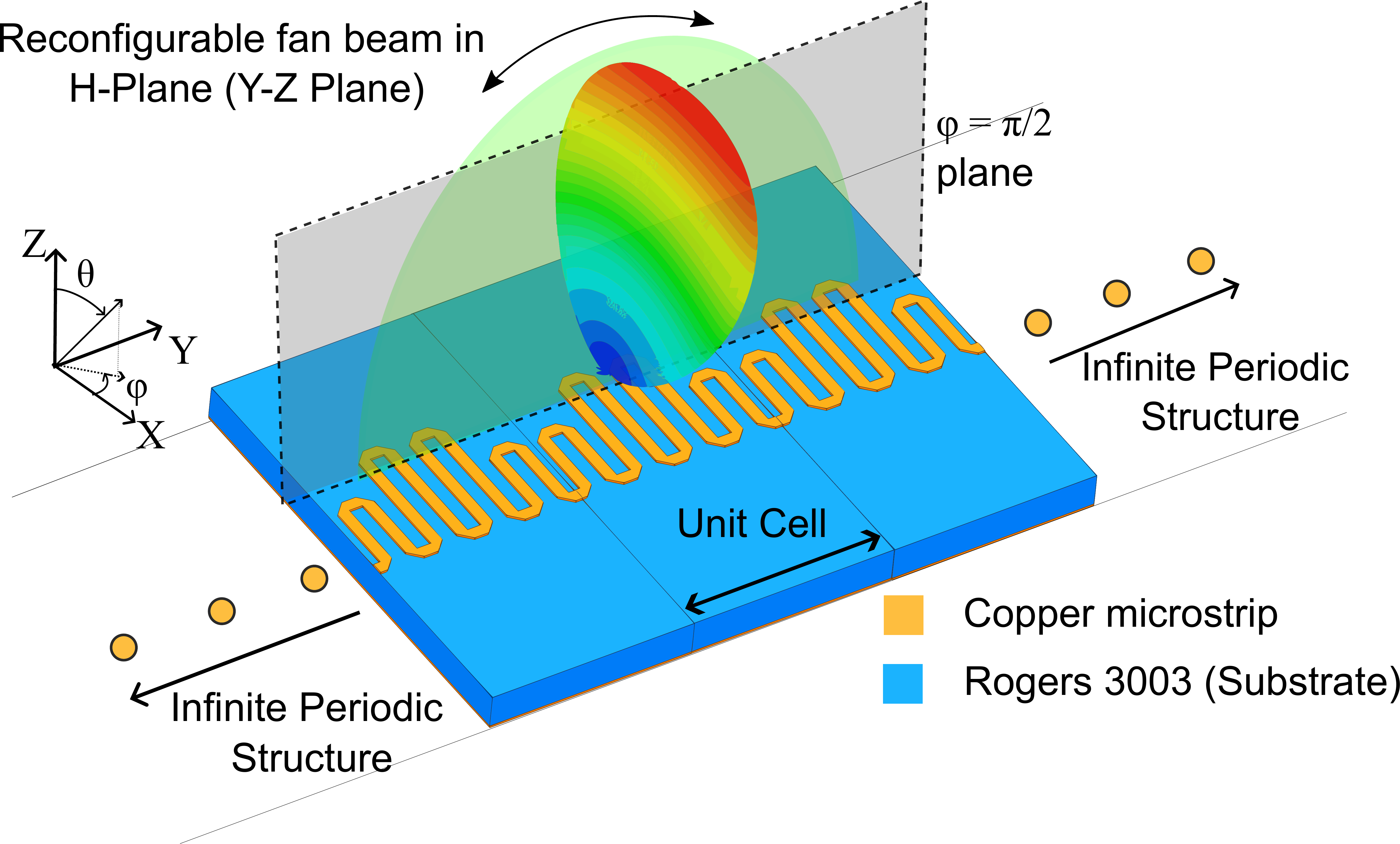}}
\caption{{~Conductor backed periodic meandering microstrip LWA.}}
\label{FullLWA}
\end{figure}

\begin{figure}[!t]\centerline{\includegraphics[width=0.35\textwidth]{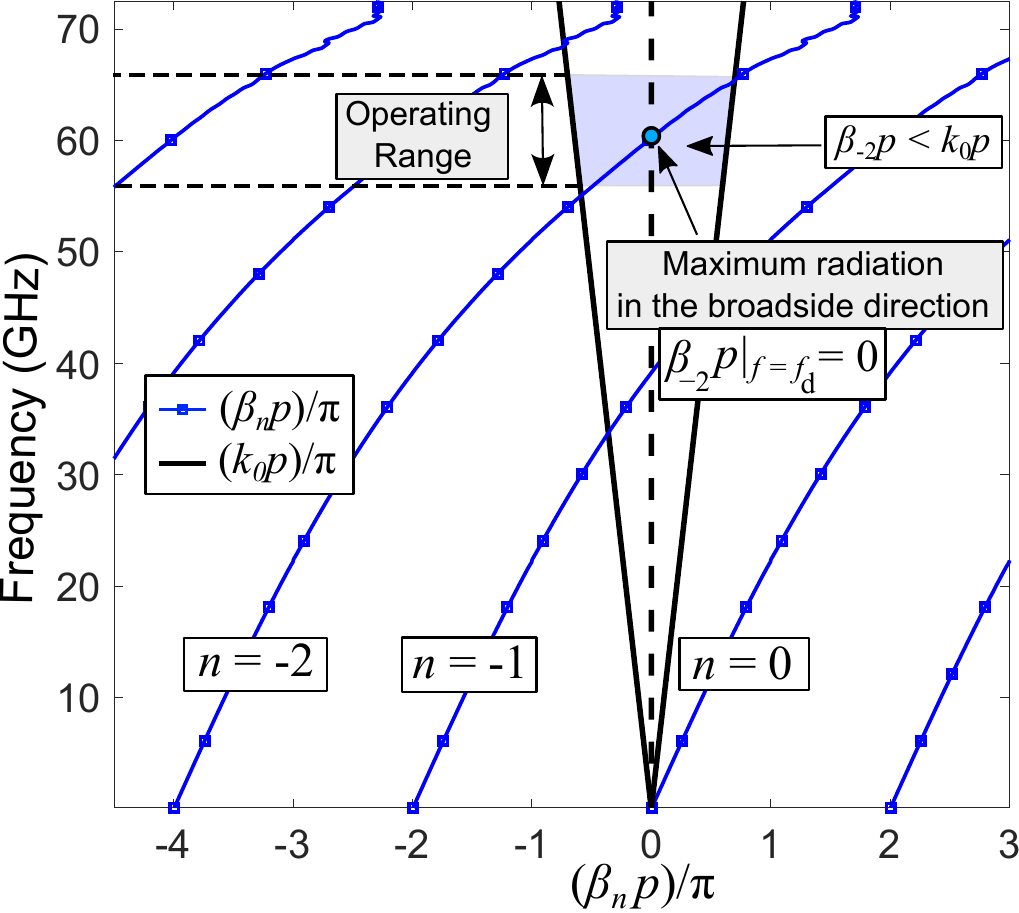}}
\caption{Brillouin diagram of the proposed leaky-wave meandering microstrip structure.}
\label{BrillouinDiagram}
\end{figure}

Fig.~\ref{FullLWA} shows the design of the proposed periodic meandering microstrip based \gls{lwa}. A \SI{0.254}{\milli\meter}-thick layer of Rogers~3003 ($\varepsilon_r = 3.0$, $\tan\delta = 0.0010$) is utilised as substrate. The structure leaks energy due to the mitred corners which act as radiating discontinuities \cite{vadher_meandering_2024, hall_microstrip_1983}. The leaky radiation due to a periodic structure with a period $p$ can be analyzed by the Bloch-Floquet theorem with the use of spatial harmonics \cite{jackson_leaky-wave_2011, collin_antenna_1969}. The phase constant corresponding to the $n^{\mathrm{th}}$ space harmonic $\beta_n$ satisfies

\begin{equation} 
\beta_{n}p = \beta_{0}p+2 n\pi\,, \label{eq1}
\end{equation}
where $n$ ranges from $-\infty$ to $+\infty$. Here $\beta_0$ is the zeroth order spatial harmonic of a periodic \gls{lwa}. 

The spatial harmonics that satisfy the condition $|\beta_n| p <|k_0|p$ contribute to the leaky-wave radiation. A guiding medium with periodic radiating discontinuities can scan the main beam from backward to forward direction. 

If most of the power from the structure is radiated, the direction of maximum radiation $\theta_{\mathrm{max}, n}$ corresponding to the $n^{\mathrm{th}}$ spatial harmonic is given by \cite{collin_antenna_1969, ishimaru_electromagnetic_2017, rahimi_higher-order_2021, harvey_periodic_1960}
\begin{equation} 
\theta_{\mathrm{max},n}=\sin^{-1}(\beta_{-n}p/k_0p)\,, \label{eq2}
\end{equation}
where $k_0$ is the free space wave number.

For the proposed structure, the $n=-2$ spatial harmonic, as shown in the Brillouin diagram in Fig.~\ref{BrillouinDiagram}, is responsible for radiation in the desired operating range (\SIrange[range-phrase=--]{57}{64}{\giga\hertz}) .

Consequently, according to equation \eqref{eq2}, when the phase difference across the unit cell ($\beta_{-n}p$) is zero, the direction of the main beam associated to the $n^{\mathrm{th}}$ spatial harmonic is broadside. Since the spatial harmonic $n=-2$ is responsible for radiation in the desired band, the phase difference across the unit cell with a period $p$ from equation \eqref{eq1} corresponds to: 

\begin{equation} 
\beta_{-2}p|_{f=f_\mathrm{d}} = 0 \rightarrow \beta_{0}p|_{f=f_\mathrm{d}} = 4 \pi\,, \label{eq4}
\end{equation}
where $f_\mathrm{d}$ is the desired frequency corresponding to the broadside radiation for $n=-2$. Equation \eqref{eq4} further implies that $\beta_{\mathrm{0, tem}}|_{f=f_\mathrm{d}}\times l = 4\pi$ \cite{vadher_meandering_2024, wood_curved_1979}. Here, $\beta_\mathrm{0, tem}$ is the phase constant of the microstrip media, $l$ denotes the total pathlength of the meandering microstrips. The pathlength of the microstrip is calculated based on the model proposed by \cite{hall_microstrip_1983}.

\begin{figure*}[!t]\centerline{\includegraphics[width=0.63\textwidth]{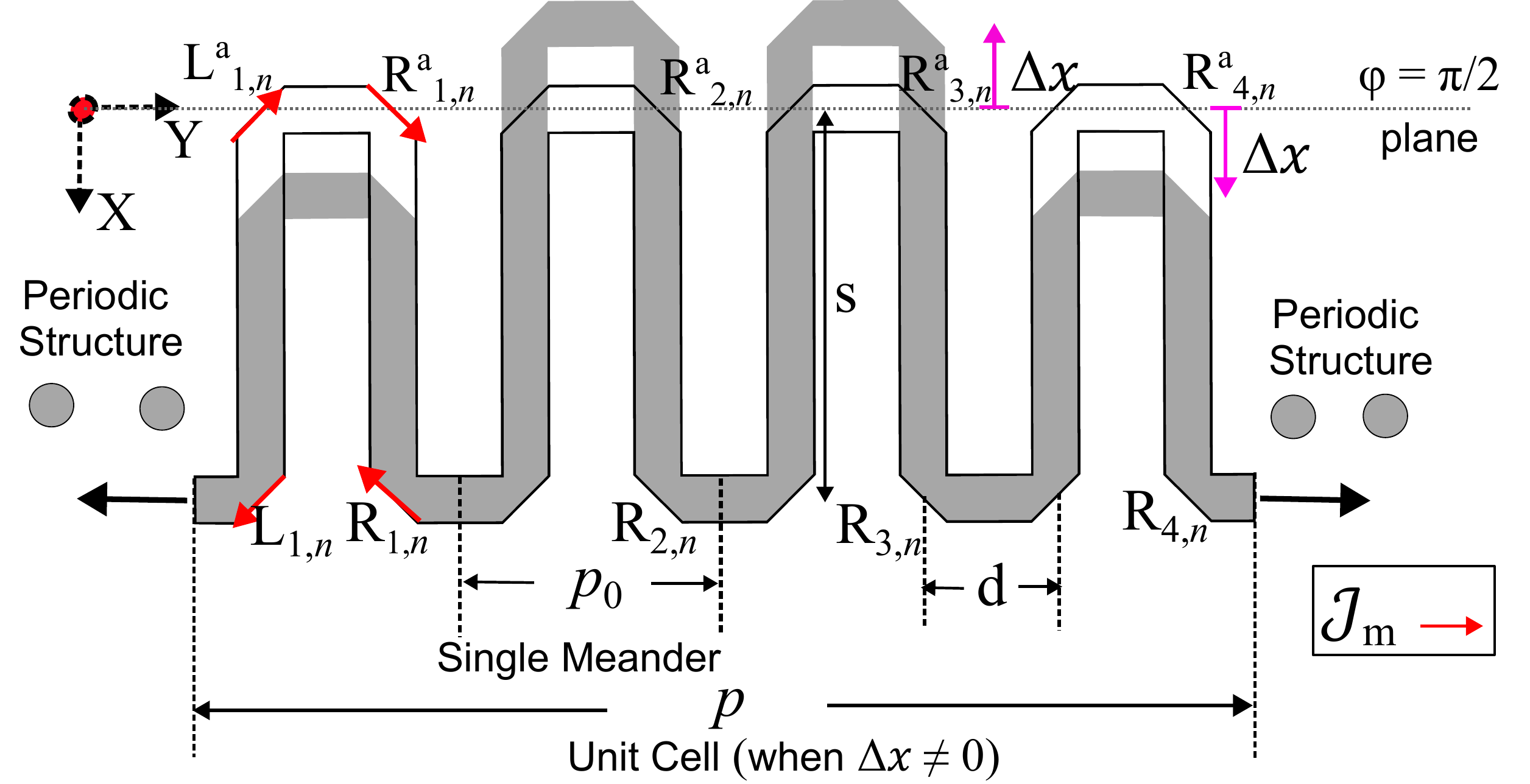}}
\caption{~Meandered microstrip guiding medium with equal meanders (shown as solid line) behaving as non-radiating slow-wave structure and unit cell formed by 4 modulated \textcolor{black}{vertical length} meanders (shown in grey) for constructive radiation leading to leaky-wave behaviour.}
\label{fig::UnitCellsCombined}
\end{figure*}

\begin{table*}[t!]
    \centering
    \caption{Geometrical parameters (Fig.~\ref{fig::UnitCellsCombined}) and corresponding phase differences at the design frequency $f_\text{d}$.}
    \renewcommand{\arraystretch}{1.5}
    \begin{tabular}{ccccc}
        \toprule
        \textbf{Segment} & \textbf{Denoted by} & \textbf{Phase diff. at $f_\text{d}$} & \multicolumn{2}{c}{\textbf{Prototyped Dimensions}} \\
        \cmidrule(r){4-5}
        & & & \textbf{(rad)} & \textbf{(mm)} \\
        \midrule
        Horizontal length of meander & $\mathrm{d}$ &  $\zeta$ & $\pi/$15 & 0.110\\
        Vertical length of meander & $\mathrm{s}$ &  $\pi/2-\zeta$ & $\pi/2-\pi/$15 & 0.868 \\
        Change in vertical length of meander & $\Delta {x}$ & $\Delta\psi$ & 0.3 & 0.158\\
        \bottomrule
    \end{tabular}
    \label{DimenTable}
\end{table*}

\begin{figure*}[!t]
    \centering
    \begin{subfigure}[t]{0.52\textwidth}
        \centering
        \includegraphics[width=\textwidth]{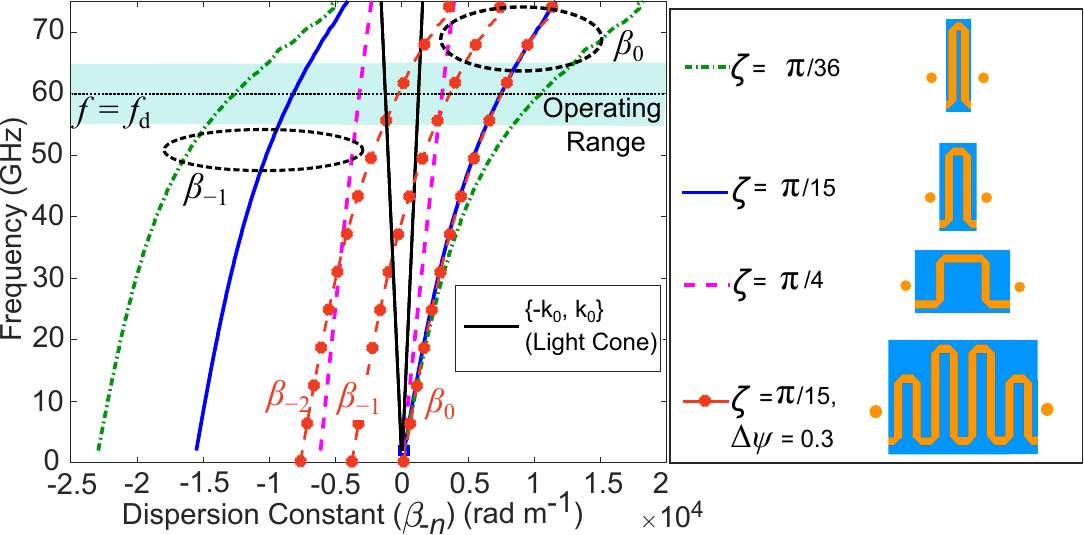}
        \caption{}
        \label{fig:DispersionControl}
    \end{subfigure}
    \hfill
    \begin{subfigure}[t]{0.44\textwidth}
        \centering
        \includegraphics[width=\textwidth]{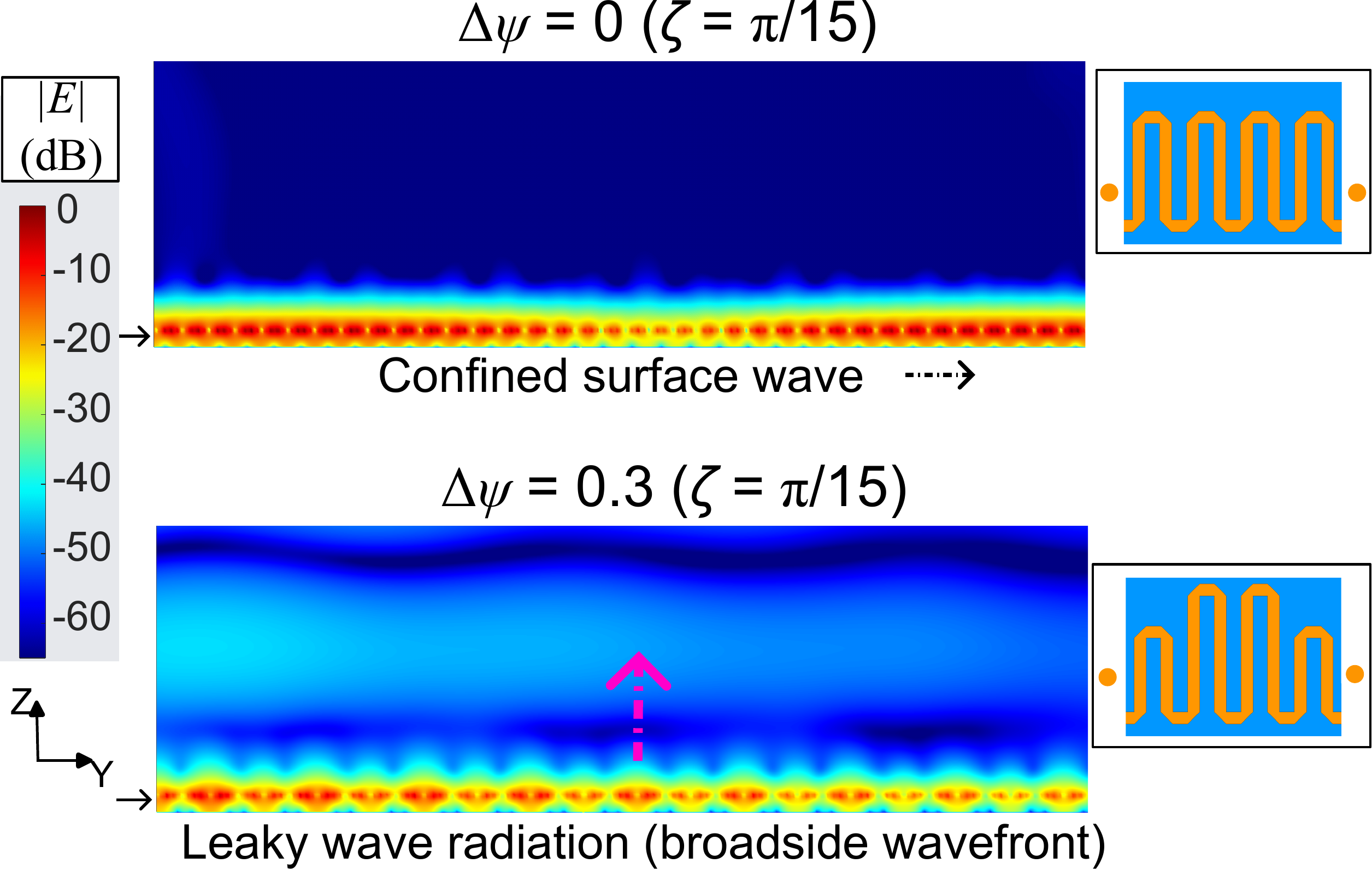}
        \caption{}
        \label{RealGainSWS_SurfaceWavef}
    \end{subfigure}
    \caption{(a)~ Dispersion diagram for structures with equal meanders ($\beta_0,\quad \beta_{-1}$ are shown), \textcolor{black}{and unit cell with periodically modulated meanders ($\beta_0, \quad \beta_{-1},\quad \beta_{-2}$ are shown) respectively.} (b)~Distribution of the absolute value of the $\bm{E}$ field in the H-Plane (Y--Z Plane). Confined surface waves at $f_\text{d}$ for a structure with equal meanders ($\Delta\psi = 0$) and leaky-wave radiation for the case $\Delta\psi = 0.3$ as the wave travels down the meandering microstrips}
    \label{fig:combined_three_side_by_side}
\end{figure*}
\begin{table*}[!t]
    \centering
    \caption{Phase shifts of $\bm{E}$ at the mitred corners corresponding to $\mathrm{R}_{i,1}$ and $\mathrm{R^a}_{i,1}$ at $f_\text{d}$ induced by the pathlength.}
    \renewcommand{\arraystretch}{1.5}
    \begin{tabular}{l *{8}{>{\centering\arraybackslash}p{0.08\linewidth}}}
        \toprule
        {Corner} & $\mathrm{R^a}_{1,1}$ & $\mathrm{R}_{1,1}$ & $\mathrm{R^a}_{2,1}$ & $\mathrm{R}_{2,1}$ & $\mathrm{R^a}_{3,1}$ & $\mathrm{R}_{3,1}$ & $\mathrm{R^a}_{4,1}$ & $\mathrm{R}_{4,1}$ \\
        {Spatial Vector Direction}& $\mathrm{(\hat{x}+\hat{y})}$ & $\mathrm{(-\hat{x}-\hat{y})}$ & $\mathrm{(\hat{x}+\hat{y})}$ & $\mathrm{(-\hat{x}-\hat{y})}$ & $\mathrm{(\hat{x}+\hat{y})}$ & $\mathrm{(-\hat{x}-\hat{y})}$ & $\mathrm{(\hat{x}+\hat{y})}$ & $\mathrm{(-\hat{x}-\hat{y})}$ \\
        \midrule
        Phase at $f_\text{d}$ & $\frac{\pi}{2} + \frac{\zeta}{2} - \Delta\psi$ & $\pi - \frac{\zeta}{2} - + 2\Delta\psi$ & $\frac{3\pi}{2} + \frac{\zeta}{2} - + \Delta\psi$ & $2\pi - \frac{\zeta}{2}$ & $\frac{5\pi}{2} + \frac{\zeta}{2} + +\Delta\psi$  & $3\pi - \frac{\zeta}{2} + +2\Delta\psi$ & $\frac{7\pi}{2} + \frac{\zeta}{2} + +\Delta\psi$ & $4\pi - \frac{\zeta}{2}$ \\
        \bottomrule
    \end{tabular}
    \label{TablePhase}
\end{table*}

\begin{table*}[!t]
    \centering
    \caption{Phase shifts of $\bm{E}$ at the mitred corners corresponding to $\mathrm{L}_{i,1}$ and $\mathrm{L^a}_{i,1}$ at $f_\text{d}$ induced by the pathlength.}
    \renewcommand{\arraystretch}{1.5}
    \begin{tabular}{l *{8}{>{\centering\arraybackslash}p{0.08\linewidth}}}
        \toprule
        {Corner} & $\mathrm{L^a}_{1,1}$ & $\mathrm{L}_{1,1}$ & $\mathrm{L^a}_{2,1}$ & $\mathrm{L}_{2,1}$ & $\mathrm{L^a}_{3,1}$ & $\mathrm{L}_{3,1}$ & $\mathrm{L^a}_{4,1}$ & $\mathrm{L}_{4,1}$ \\
        {Spatial Vector Direction}& $\mathrm{(-\hat{x}+\hat{y})}$ & $\mathrm{(\hat{x}-\hat{y})}$ & $\mathrm{(-\hat{x}+\hat{y})}$ & $\mathrm{(\hat{x}-\hat{y})}$ & $\mathrm{(-\hat{x}+\hat{y})}$ & $\mathrm{(\hat{x}-\hat{y})}$ & $\mathrm{(-\hat{x}+\hat{y})}$ & $\mathrm{(\hat{x}-\hat{y})}$ \\
        \midrule
        Phase at $f_\text{d}$ & $\frac{\pi}{2} - \frac{\zeta}{2} - \Delta\psi$ & $\frac{\zeta}{2}$ & $\frac{3\pi}{2} - \frac{\zeta}{2} - +\Delta\psi$ & $\pi + \frac{\zeta}{2} - +2\Delta\psi$ & $\frac{5\pi}{2} - \frac{\zeta}{2} ++ \Delta\psi$  & $2\pi + \frac{\zeta}{2}$ & $\frac{7\pi}{2} - \frac{\zeta}{2} + +\Delta\psi$ & $3\pi + \frac{\zeta}{2} + +2\Delta\psi$  \\
        \bottomrule
    \end{tabular}
    \label{TablePhaseL}
\end{table*}

\subsection{Non-radiating Guiding Medium with Dispersion Control}
A meandering microstrip structure with equal meanders is shown in Fig.~\ref{fig::UnitCellsCombined} (depicted as solid boundary line). The phase contributions corresponding to the horizontal and vertical intervals of the meander are listed in Table~\ref{DimenTable}. A guiding medium based on the meandering microstrip allows to finely control the path of the traveling wave by altering the length of the intervals between the corners of the meander~\cite{tianConformalPropagationNearomnidirectional2020,tian_implant--implant_2023 }.

The proposed \gls{sws} is designed such that a single meander results in a phase difference of $\pi$ at $f_\text{d}$. The rate of change of the phase constant ($\beta$) with frequency (i.e., the dispersion rate) can be increased by {shrinking} the period $p_0$ of the single meander. The period $p_0$ can be reduced---while maintaining the same electrical length of the meander---by decreasing the horizontal interval and concurrently increasing the vertical interval of the meander by the same amount. 
 
Consequently, the impact of the variation of the horizontal segment (denoted by $\zeta$) on $\beta$ is shown in Fig.\ref{fig:DispersionControl} indicating a high control over the dispersion rate of the meandering microstrip guiding medium. In the desired operating range of \SI{57}{\GHz} to \SI{64}{\GHz}, the dispersion constant of the structure lies outside the light cone, resulting in a confined surface wave within the guiding medium without any leaky radiation. This is illustrated in Fig.~\ref{RealGainSWS_SurfaceWavef}, which shows the magnitude of the E-field obtained from the full-wave simulation of 72 cascaded meanders (equivalent to the final length of the \gls{lwa}), all having equal lengths with the geometrical parameters $\zeta = \pi/15$ and $\Delta\psi=0$ at $f_\text{d}$.

\begin{figure}[!t]\centerline{\includegraphics[width=0.33\textwidth]{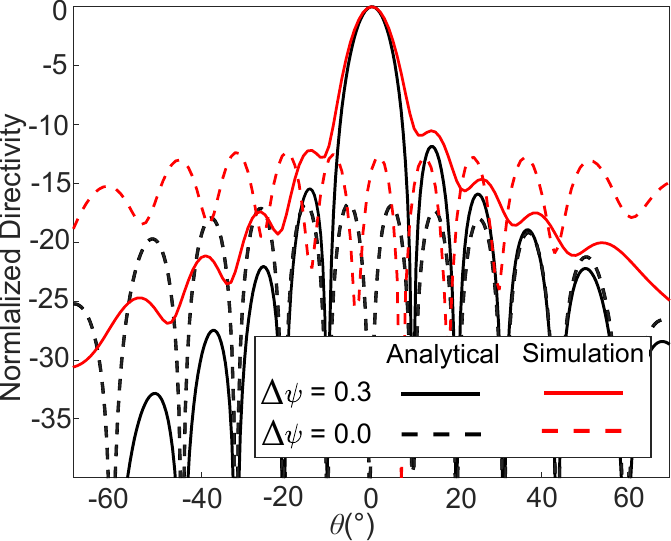}}
\caption{~\textcolor{black}{Changes in the directivity observed through analytical formulation and simulation of the proposed structure at $f_\text{d}$ (in the broadside radiation) for two cases of $\Delta\psi$ (for $\zeta = \pi/15$).}}
\label{fig:AnalyticalvsSimulation}
\end{figure}

\subsection{Meandered LWA with Periodically Modulated Vertical Length}
To obtain a periodic unit cell, four meanders designed in the previous section are cascaded in series [Fig.\ref{fig::UnitCellsCombined}]. It is possible to modify the vertical length of the meanders without changing the total pathlength of the unit cell ($4\pi$ for four meanders at $f_{\text{d}}$) to obtain a radiation in the far-field at $f_\text{d}$. This is achieved by decreasing the vertical length of the first meander by ${\Delta x}$, and increasing the length of the second meander by the same quantity. A length ${\Delta x}$ results in a phase delay of $\Delta\psi$ at $f_\text{d}$ (Table.~\ref{DimenTable}). The third and fourth meanders are modified accordingly to obtain a symmetric unit cell. The resultant unit cell is depicted in grey color in Fig.\ref{fig::UnitCellsCombined}.

\textcolor{black}{The dispersion diagram is plotted in Fig.~\ref{fig:DispersionControl} where three spatial harmonics ($\beta_0, \quad \beta_{-1},\quad \beta_{-2}$) for different unit cells are shown. Two observations can be made in the diagram: (1.) for the proposed unit cell with unequal meanders ($\zeta = \pi/15$ and $\Delta\psi=0.3$) and the one with equal meanders ($\zeta = \pi/15$ and $\Delta\psi = 0$), the $\beta_0$ curves overlap. However, since the period of the proposed unit cell consists of 4 meanders, this results in two more closely spaced spatial harmonics ($\beta_{-1}$ and $\beta{-2}$) (2.) due to $|\beta_{-2}| < k_0$ in the operating range for the proposed unit cell, there will be leaky radiation. Further, at $f_\mathrm{d}$, since $\beta_{-2}p = 0$ [described in Equation~\eqref{eq4}], the radiation is in the broadside direction. The radiation diagram is shown in the Fig.~\ref{RealGainSWS_SurfaceWavef} ($\zeta = \pi/15$ and $\Delta\psi=0.3$).}

\subsection{\textcolor{black}{Far-field Calculations Based on Array Theory for the Proposed LWA}}
The radiation resulting from a microstrip based \gls{lwa} can be also be analyzed in terms of residual magnetic current densities (${{\mathcal{J}_\mathrm{m}}}$) at the corners \cite{wood_curved_1979, harrington_fundamental_2001}. The  residual currents at each corner are shown in  Fig.~\ref{fig::UnitCellsCombined}. To achieve a constructive addition of the electric fields from the residual currents, the vertical length of the meanders is periodically varied. 

For the proposed unit cell, the corners are denominated by $\mathrm{L}_{{{i, n}}}$, $\mathrm{L^{\text{a}}}_{{{i, n}}}$, $\mathrm{R^{\text{a}}}_{{{i, n}}}$, and $\mathrm{R}_{{{i, n}}}$, where $i$ $\in$ ([$1, \mathrm{4}$]) indicates the $i^\mathrm{th}$ meander within the unit cell, and $n \in [1, \mathrm{N}]$ corresponds to the $n^\mathrm{th}$ unit cell, $\mathrm{N}$ being the total number of unit cells (Fig.~\ref{fig::UnitCellsCombined}). 

The microstrip line intervals result in a phase delay between the corners. The phases corresponding to the corners of the first unit cell are indicated in Table~\ref{TablePhase} and Table~\ref{TablePhaseL}.

Since equation \eqref{eq4} is satisfied at $f_\text{d}$, the radiation occurs in the broadside direction. Subsequently, the ratio of the total electric field generated by N such unit cells is obtained as the summation of the electric fields for the unit cell \cite{james_microstrip_1977} shown in Fig.~\ref{fig::UnitCellsCombined} generated by the ${{\mathcal{J}_\mathrm{m}}}$ oriented towards the direction $\mathrm{(+\hat{x}+\hat{y})}$ is: 

\begin{equation}
\frac{E^{\text{tot}}_{\mathrm{{R^a}}_{2, n}|f=f_\mathrm{d}}}{E^{\text{tot}}_{ \mathrm{{R^a}}_{1, n}|f=f_\mathrm{d}}} = \frac{\sum_{{n}=1}^\mathrm{N} e^{-j({\frac{3\pi}{2} + \frac{\zeta}{2} - \Delta\psi}+4(n-1)\pi)}}{\sum_{{n}=1}^\mathrm{N}e^{-j({{\frac{\pi}{2} + \frac{\zeta}{2} - \Delta\psi}}+4(n-1)\pi)}} = -1
\label{Ra2Ra1}
\end{equation}

\begin{equation}
\frac{E^{\text{tot}}_{\mathrm{{R^a}}_{4, n}|f=f_\mathrm{d}}}{E^{\text{tot}}_{ \mathrm{{R^a}}_{3, n}|f=f_\mathrm{d}}} = \frac{\sum_{{n}=1}^\mathrm{N}e^{-j({\frac{7\pi}{2} + \frac{\zeta}{2} + \Delta\psi}+4(n-1)\pi)}}{\sum_{{n}=1}^\mathrm{N}e^{-j({{\frac{5\pi}{2} + \frac{\zeta}{2} + \Delta\psi}}+4(n-1)\pi)}} = -1
\label{Ra4Ra3}
\end{equation}

The ratio of the total electric field generated by the summation of the contributions ${{\mathcal{J}_\mathrm{m}}}$ oriented towards the direction $\mathrm{(-\hat{x}-\hat{y})}$ at $f_\text{d}$ corresponds to: 

\begin{equation}
\frac{E^{\text{tot}}_{\mathrm{{R}}_{2, n}|f=f_\text{d}}}{E^{\text{tot}}_{ \mathrm{{R}}_{1, n}|f=f_\text{d}}} = \frac{\sum_{{n}=1}^\mathrm{N}e^{-j({2\pi - \frac{\zeta}{2}}+4(n-1)\pi)}}{\sum_{{n}=1}^\mathrm{N}e^{-j({{\pi - \frac{\zeta}{2} - 2\Delta\psi}+4(n-1)\pi})}} = -e^{-2j\Delta\psi}
\label{R2R1}
\end{equation}

\begin{equation}
\frac{E^{\text{tot}}_{\mathrm{{R}}_{4, n}|f=f_\mathrm{d}}}{E^{\text{tot}}_{ \mathrm{{R}}_{3, n}|f=f_\mathrm{d}}} = \frac{\sum_{{n}=1}^\mathrm{N}e^{-j({4\pi - \frac{\zeta}{2}}+4(n-1)\pi)}}{\sum_{{n}=1}^\mathrm{N}e^{-j({{3\pi - \frac{\zeta}{2} + 2\Delta\psi}}+4(n-1)\pi)}} = -e^{2j\Delta\psi}
\label{R4R3}
\end{equation}

Equations \eqref{Ra2Ra1} and \eqref{Ra4Ra3} show that the electric fields at the corners $\mathrm{R^a}_{i, n}$ are opposite to each other, while the electric field vectors at $\mathrm{R}_{i, n}$ ($i\in[1, 4]$) contribute to the far-field. By repeating the same analysis for the corners $\mathrm{L}_{i, n}$ and $\mathrm{L^\text{a}}_{i, n}$ (the corresponding phases are indicated in Table.~\ref{TablePhaseL}) it is possible to derive the same conclusion.

Note that if $\Delta\psi=0$, the resultant electric field is zero, reducing the structure to the non-radiating \gls{sws} shown in Fig.~\ref{fig::UnitCellsCombined} (equal meanders, depicted as solid line).

\begin{figure*}[!t]
    \centering
    \begin{subfigure}[b]{0.325\textwidth}
        \centering
            \includegraphics[width=\textwidth]{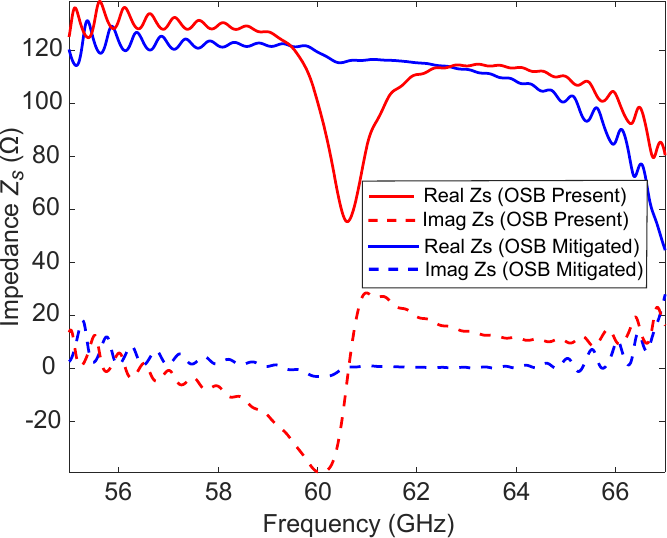}
        \caption{}
        \label{fig:BlochImpedance}
    \end{subfigure}
    \hfill
    \begin{subfigure}[b]{0.325\textwidth}
        \centering
        \includegraphics[width=\textwidth]{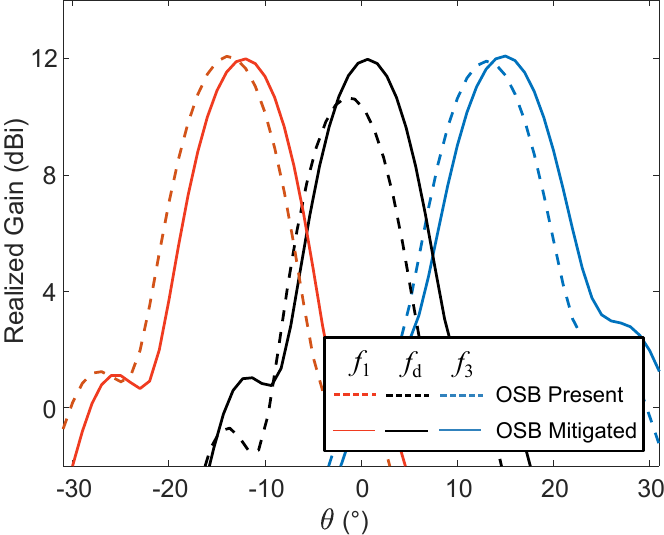}
        \caption{}
        \label{FarFieldonlyLWA}
    \end{subfigure}
    \hfill
    \begin{subfigure}[b]{0.285\textwidth}
        \centering
        \includegraphics[width=\textwidth]{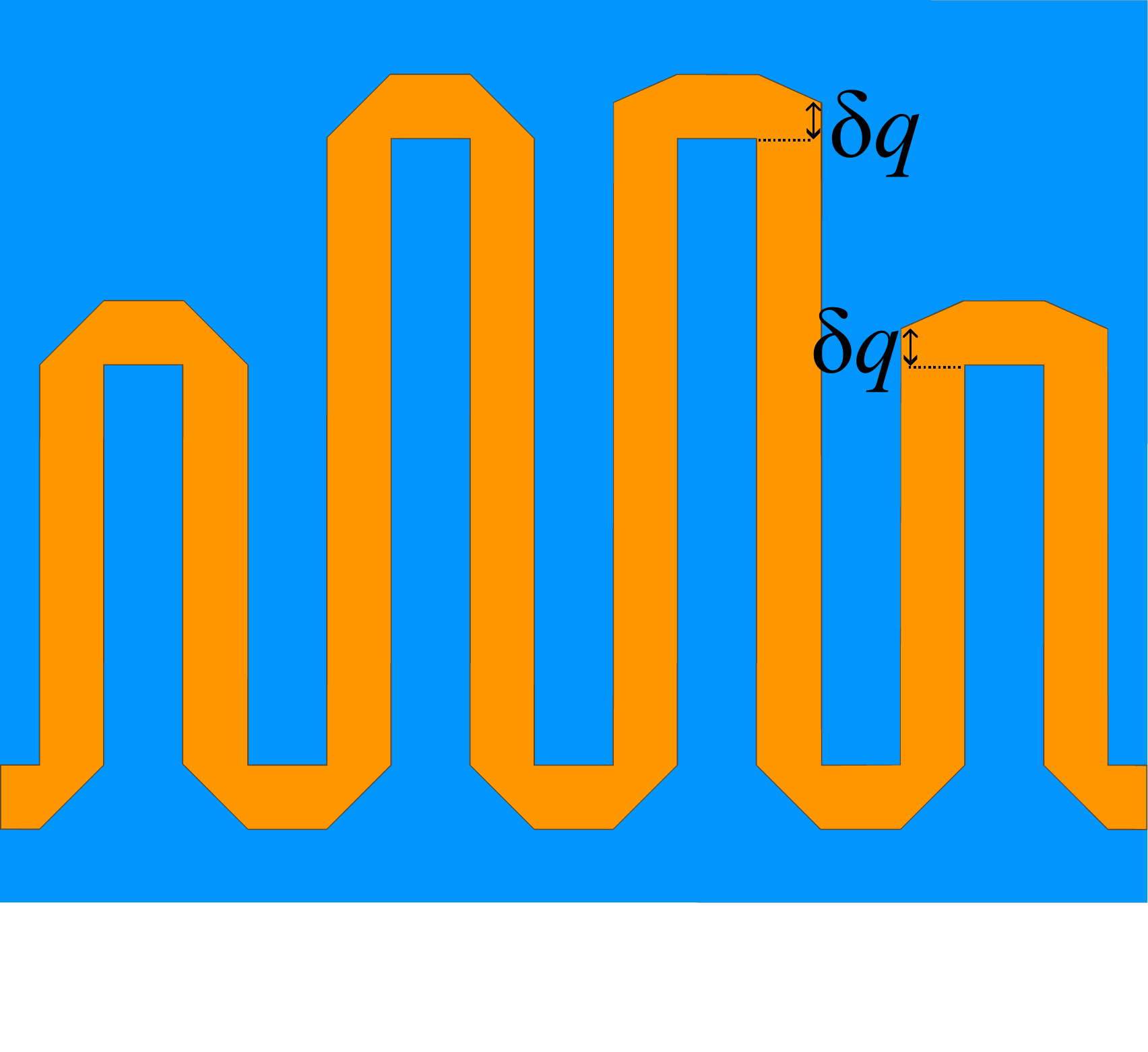}
        \caption{}
        \label{fig:OSBRemoval}
    \end{subfigure}
    \caption{(a)~Bloch impedance $Z_\mathrm{s}$ vs frequency, (b)~Impact on the realized gain before and after removal of OSB at the frequency $f_\text{d}$ in the broadside direction. Here, $f_1 = f_\mathrm{d} -$~\SI{1}{\giga\hertz} and $f_1 = f_\mathrm{d} + $~\SI{1}{\giga\hertz}. (c)~Removal of \textcolor{black}{the} OSB by changing the angle of mitred corner for the third and fourth meander of the unit cell.}
    \label{}
\end{figure*}

\begin{figure}[!t]\centerline{\includegraphics[width=0.32\textwidth]{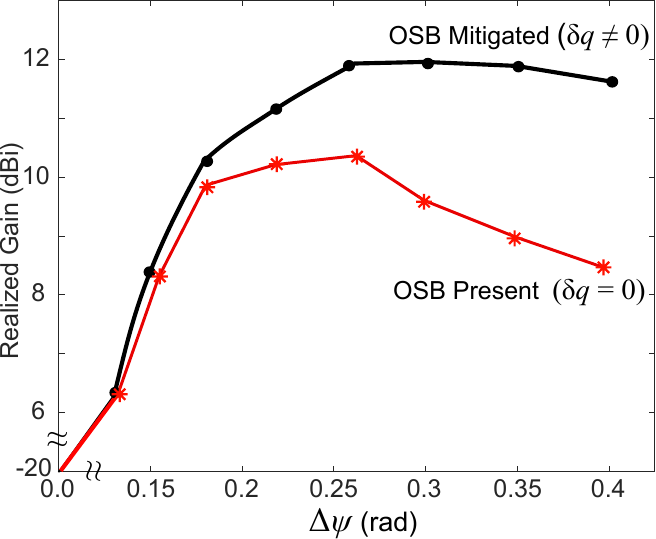}}
\caption{Realized gain at $f_\text{d}$ (in the broadside radiation) as a function of $\Delta\psi$ (for $\zeta = \pi/15$).}
\label{fig:RealGainSWS_JustBroadside}
\end{figure}

\textcolor{black}{An analytical model \cite{sacco_analysis_2021} based on the equations \eqref{Ra2Ra1},\eqref{Ra4Ra3}, \eqref{R2R1}, and \eqref{R4R3} is designed to verify the effectiveness and the results are compared with full-wave-simulations as shown in Fig.~\ref{fig:AnalyticalvsSimulation} at $f_\text{d}$.} As can be observed, when $\Delta\psi=0$, all the meanders are of equal size leading to no radiation. When the meanders become unequal with $\Delta\psi=$ 0.3, there is radiation. The above model adopts simplifying assumptions, treating radiation sources as magnetic line currents at the corners, as proposed by \cite{hall_microstrip_1983, sacco_analysis_2021}. Additionally, it assumes that the attenuation of the traveling wave down the line is minimal and that the bends are matched, resulting in no reflections. Although this approach loses information about mutual coupling, the effect of \textcolor{black}{the} \gls{osb} and substrate surface waves, it offers closed-form design solutions to guide the design of the proposed meandering-microstrip \gls{lwa} for the desirable radiation characteristics at the frequency of broadside radiation \cite{james_microstrip_1986}.

\begin{figure*}[!t]\centerline{\includegraphics[width=0.75\textwidth]{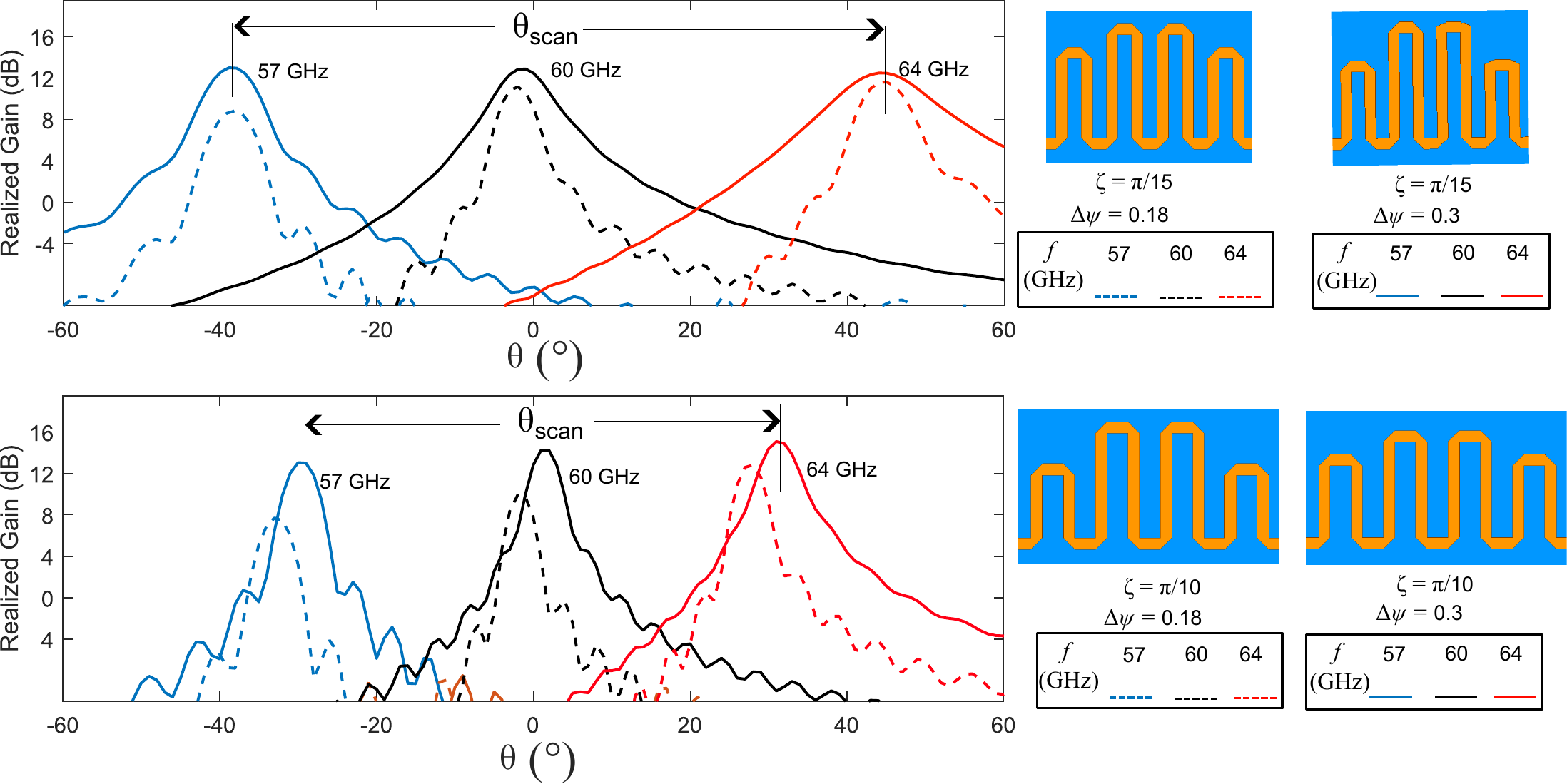}}
\caption{\textcolor{black}{Illustration of effective control of scanning rate (${S_\mathrm{m}}$) and gain of the \gls{lwa} with the depicted unit cells. For the fabricated prototype $\zeta = \pi/15$ and $\Delta\psi = 0.3$ is used.}}
\label{fig: thetaH0_2094_thetaDDiff}
\end{figure*}

\begin{figure*}[!t]
    \centering
    \begin{subfigure}[b]{0.32\textwidth}
        \centering
        \includegraphics[width=\textwidth]{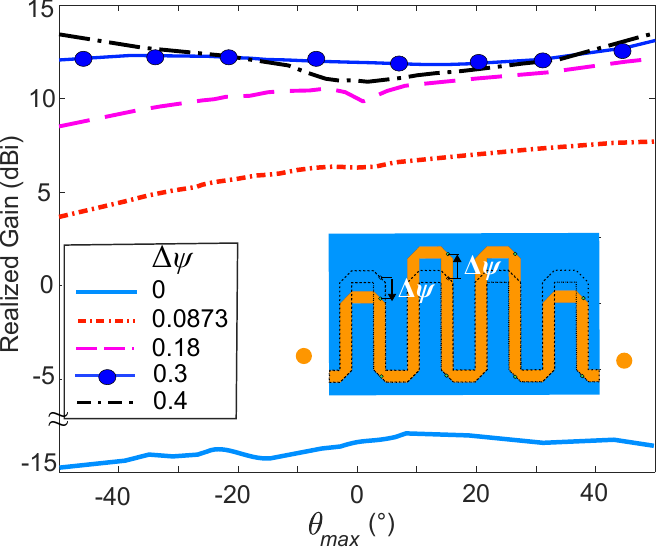}
        \caption{}
        \label{fig:RGvsThetaMax}
    \end{subfigure}
    \hfill
    \begin{subfigure}[b]{0.32\textwidth}
        \centering
        \includegraphics[width=\textwidth]{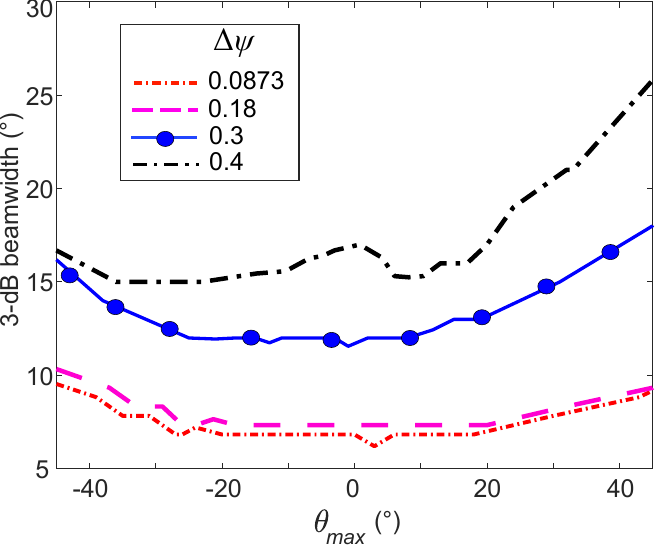}
        \caption{}
        \label{fig:3dBvsThetaMax}
    \end{subfigure}
    \hfill
        \begin{subfigure}[b]{0.326\textwidth}
        \centering
        \includegraphics[width=\textwidth]{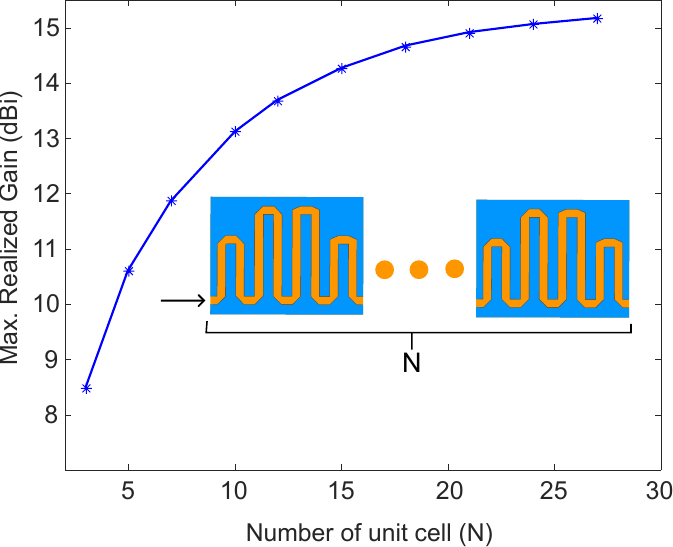}
        \caption{}
        \label{RGvsN}
    \end{subfigure}
    \caption{(a)~Maximum realized gain following the removal of the \gls{osb} in the H-plane ($\phi = \pi/2$) for different $\Delta\psi$ as a function of the beam pointing direction, (b)~3-\si{dB} beamwidth of scanning beam vs frequency for different values of $\Delta\psi$ at $f_\text{d}$. (c)~ Maximum broadside realized gain at $f_\text{d}$ as a function of the number of unit cells for $\Delta\psi =$ 0.3 and $\zeta = \pi$/15. }
    \label{fig:combined_last_two}
\end{figure*}

\subsection{OSB Mitigation}
A periodic structure is characterized by the presence of stop-bands, that result in a very low transmission of the wave through the structure and in high energy being reflected back to the source \cite{jackson_leaky-wave_2011}. \textcolor{black}{The} \gls{osb} occurring inside the radiation zone limits the scanning near the broadside direction \cite{jackson_leaky-wave_2011, jackson_classical_nodate} and needs to be mitigated.  


 The Bloch impedance {$Z_\mathrm{s}$} can be used to characterize the presence of \textcolor{black}{the} \gls{osb} \cite{caloz_electromagnetic_2005}. The $Z_\mathrm{s}$ can be calculated from the ABCD-parameters obtained from the S-parameters computed in full-wave simulations \cite{pozar_microwave_2011}:

\begin{equation}
Z_\mathrm{s}= \frac{-2\times\mathrm{B}}{(\mathrm{A}-\mathrm{D}-\sqrt{(\mathrm{A}+\mathrm{D})^2-4})} \,.\label{zbloch}
\end{equation}

Fig.~\ref{fig:BlochImpedance} indicates that there is a resonance in the band around $f_{\mathrm{d}}$ contributing to \textcolor{black}{the} \gls{osb}.
The impact of \textcolor{black}{the} \gls{osb} on the antenna realized gain is shown in Fig.~\ref{FarFieldonlyLWA}. There is a severe reduction in gain in the broadside direction if \textcolor{black}{the} \gls{osb} is not mitigated. 

Asymmetry can be introduced in the unit cell to mitigate \textcolor{black}{the} \gls{osb} as shown in \cite{wang_periodic_2022, vadher_meandering_2024, otto_transversal_2014}. Changing the angles of corners of the final two meanders as shown in Fig.\ref{fig:OSBRemoval} introduces a small phase shift between the first couple and the second couple meanders to avoid strong resonance and large reflection in the band around $f_\mathrm{d}$. \textcolor{black}{The} \gls{osb} mitigation can be observed from the Bloch impedance $Z_\mathrm{s}$ plot with frequency [see Fig. \ref{fig:BlochImpedance}]. For $\zeta = \pi/$15 and $\Delta\psi~=~$0.3, a value of $\delta q$ equal to \SI{0.05}{\milli\meter} is selected. The increase in the realized gain can be observed at the broadside direction in Fig.~\ref{fig:BlochImpedance}. Additionally, since the corners  $\mathrm{L^{\text{a}}}_{{{i, n}}}$ and $\mathrm{R^{\text{a}}}_{{{i, n}}}$ do not contribute to the radiation at $f_\mathrm{d}$, they have been altered.

Following the mitigation of \textcolor{black}{the} \gls{osb}, the maximum realized gain obtained at the broadside frequency as a function of $\Delta \psi$ is depicted in Fig.\ref{fig:RealGainSWS_JustBroadside} with $\mathrm{N} =$ 30 unit cells for $\zeta = \pi/$15.

\subsection{Impact of Variation of $\zeta$ and $\Delta\psi$ in the Operating Band}
Full-wave simulations with $\mathrm{N} =$ 30 unit cells have been performed to quantify the impact of $\Delta \psi$ on the \gls{lwa} far-field in the band of operation (\SIrange[range-phrase=--,range-units=single]{57}{64}{\giga\hertz}) after the mitigation of \textcolor{black}{the} \gls{osb}. 

\textcolor{black}{Fig.~\ref{fig: thetaH0_2094_thetaDDiff} shows the effective control of gain and $S_\mathrm{m}$. For $\zeta~=~\pi/15$ the scanning range obtained is \SIrange{-40}{43}{\degree} while for $\zeta~=~\pi/10$, the scanning range obtained is \SIrange{-30}{38}{\degree}. The gain can then be varied with the modulation in the vertical length of meanders. 
Additionally, Fig.~\ref{fig:RGvsThetaMax} shows the change in realized gain for different values of $\Delta\psi$ with $\zeta = \pi/$15. Figs.~ \ref{fig:RealGainSWS_JustBroadside} and \ref{fig:RGvsThetaMax} show that the realized gain initially increases with an increase in $\Delta\psi$, but then levels off after certain values of $\Delta\psi$ ($=$ 0.3 in the current case). This indicates that the unit cell becomes a more effective radiator as $\Delta\psi$ increases up to this point. As the radiation from each unit cell increases, the effective aperture of the \gls{lwa} reduces, resulting in larger 3-$\mathrm{dB}$ beamwidth as indicated in Fig.\ref{fig:3dBvsThetaMax}.} 

As a result, the performance of the designed \gls{lwa} can be controlled through $\zeta$ (acting on $S_\textrm{m}$ and the scanning range) and $\Delta\psi$ (impacting the realized gain and 3-$\mathrm{dB}$ beamwidth). For the current prototype, the chosen parameters are listed in {Table \ref{DimenTable}} which result in a realized gain of \SI{10}{\dB} in the desired operating frequency range and angular scanning of \SI{80}{\degree} symmetrically through broadside and across the operating band. 

\subsection{Impact of the Number of Unit Cells}
To maximize the radiation efficiency and radiate most of the energy from the leaky-wave structure, while keeping it compact, determining the number of unit cells to be included in the final design is of the utmost importance. Fig.~\ref{RGvsN} shows the change in gain with the increase in the number of unit cells. After $\textrm{N}=$ 18 unit cells, the change in gain is minimal.

Further increasing the number of unit cells would result in a reduction of the radiation efficiency as well as in an increase of the antenna dimensions. It is worth noting that the performance of a longer antenna is more likely to be impacted by the bending. Hence $\textrm{N}=$ 18 unit cells is chosen as the optimal length.

\section{Prototype and Measurements for the Proposed Antenna}\label{SectionPrototype}

\captionsetup[table]{name=TABLE,labelsep=newline,textfont=sc, justification=centering}
\renewcommand{\arraystretch}{1.3}

\subsection{Design of the Patch to Terminate the \Gls{lwa}}
\begin{figure}[!t]\centerline{\includegraphics[width=0.33\textwidth]{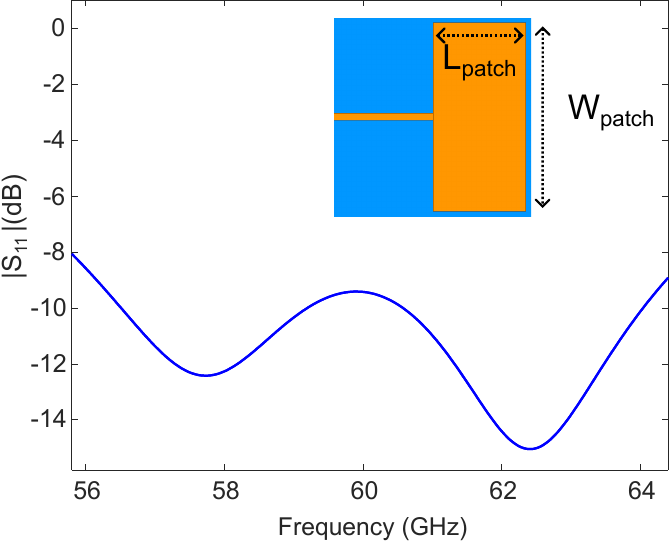}}
\caption{\textcolor{black}{$\mathrm{|S_{11}|}$ of the patch designed to terminate the LWA.}}
\label{TermPatch}
\end{figure}

\begin{figure*}[!t]
    \centering
    \begin{subfigure}[t]{0.32\textwidth}
        \centering
        \includegraphics[width=\textwidth]{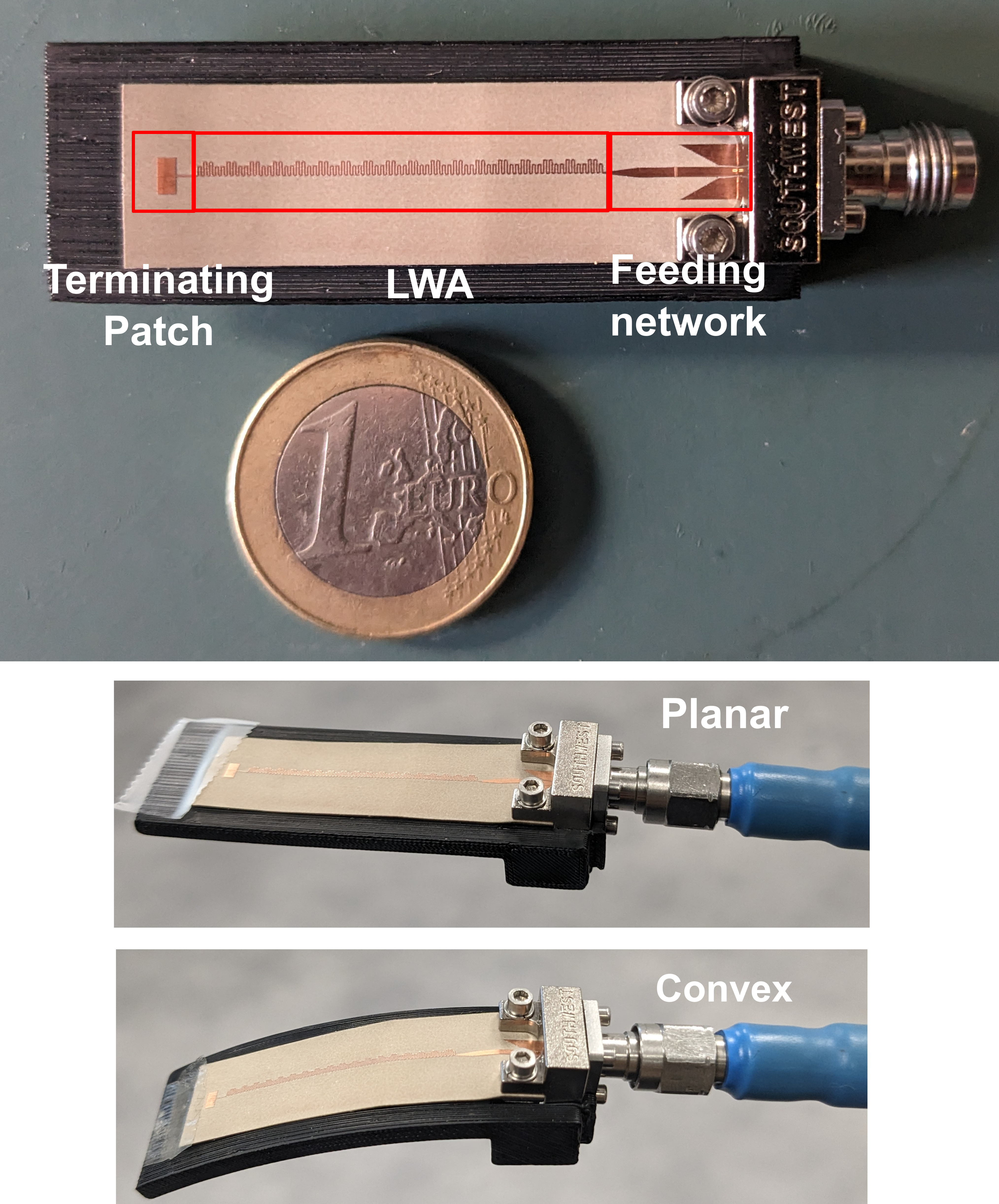}
        \caption{}
        \label{PhotoPrototype}
    \end{subfigure}
    \begin{subfigure}[t]{0.43\textwidth}
        \centering
        \includegraphics[width=\textwidth]{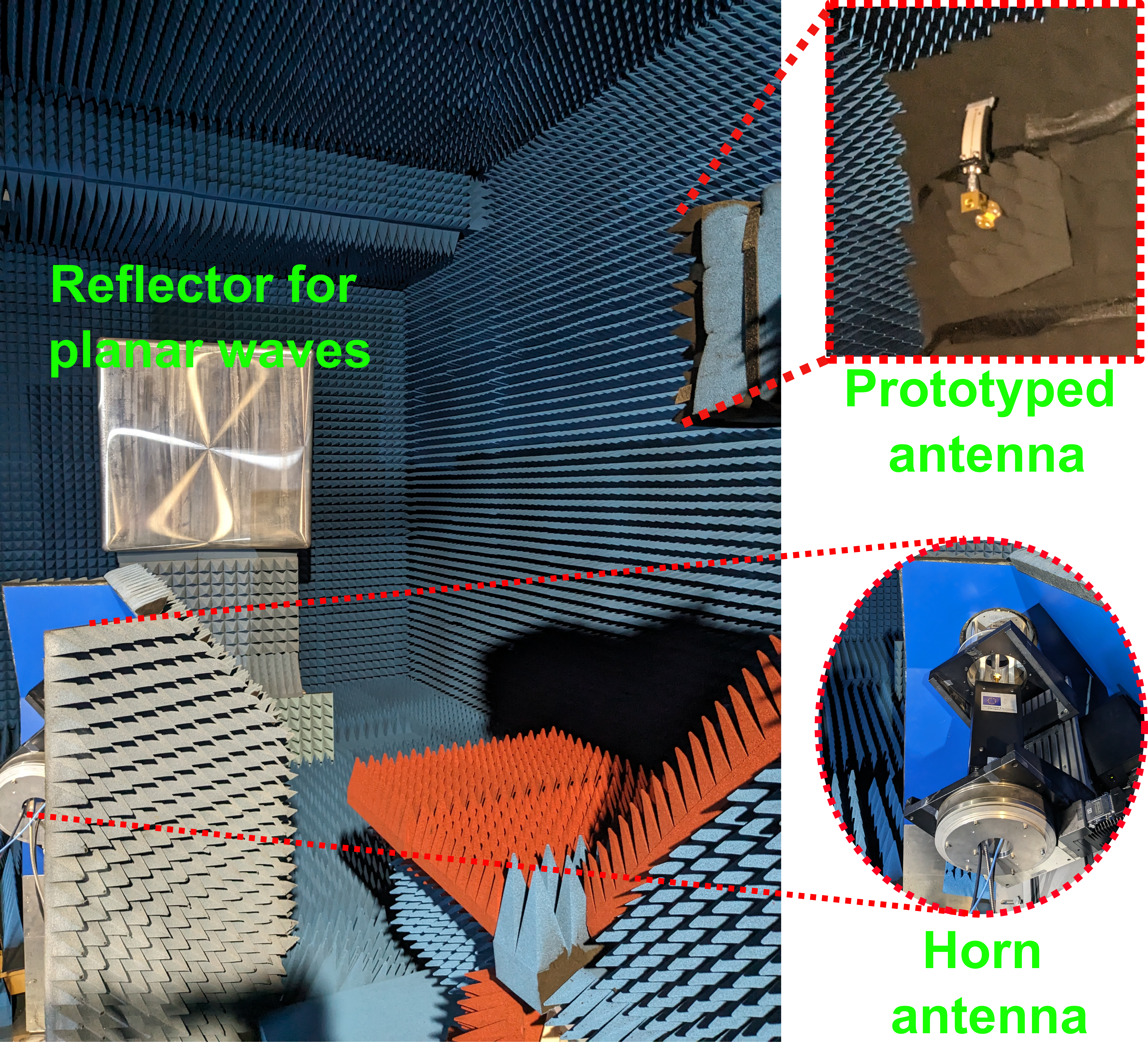}
        \caption{}
        \label{PhotoPrototypeMeasure}
    \end{subfigure}
    \caption{\textcolor{black}{(a)~Fabricated prototype of the LWA and the two states of measurements (Planar and convex) (b)~CATR measurement setup for the prototype.}}
\end{figure*}

Since, only one port of the \gls{lwa} is used to feed the antenna, the second port of the \gls{lwa} is terminated with a matched patch to avoid the use of second connector \textcolor{black}{\cite{mishra_high_2020}}, that would increase the overall dimensions of the prototype and the complexity of the measurement setup in the anechoic chamber (inset figure in Fig.~\ref{PhotoPrototypeMeasure}). The patch is designed to be well matched for the operating frequency to terminate the \gls{lwa} without any small reflections. The dimensions for the patch are listed in the Table~\ref{FeedTerminatingPatchDimensions}, while the $\mathrm{|S_{11}|}$ is shown in Fig.~\ref{TermPatch}. \textcolor{black}{As can be observed in the Fig.~\ref{fig: Efieldtermwithpatch}, most of the energy has already been radiated from the \gls{lwa} with 18 unit cells (from Fig.~\ref{RGvsN}) before reaching the patch. As a consequence, the radiation from the patch is negligible and does not affect the radiation from the \gls{lwa}.}

\subsection{Design of the Feed}
\begin{table}[]
\begin{center}
\caption{Dimensions of the feed and the terminating patch.}
\label{FeedTerminatingPatchDimensions}
\begin{tabular}{>{\centering\arraybackslash}p{0.35\linewidth} >{\centering\arraybackslash}p{0.3\linewidth}}
\toprule
\textbf{Dimensions} & \textbf{Physical dim. (mm)} \\
\midrule
$\mathrm{Co_{SL}}$ & 2.8 \\
$\mathrm{Co_{T}}$ & 1.5 \\
$\mathrm{Co_{G}}$ & 2.47 \\
$\mathrm{Co_{L}}$ & 1.75 \\
$\mathrm{t_{50}}$ & 0.69 \\
$\mathrm{Tx}$ & 2.5 \\
$\mathrm{W_{patch}}$ & 2.55 \\
$\mathrm{L_{patch}}$ & 1.23 \\
\bottomrule
\end{tabular}
\end{center}
\end{table}

\begin{figure}[!t]\centerline{\includegraphics[width=0.50\textwidth]{Figures/NumberofUnitCellsBothplane.pdf}}
\caption{\textcolor{black}{Electric field in the X--Y and Y--Z plane of the proposed LWA terminated with patch.}}
\label{fig: Efieldtermwithpatch}
\end{figure}

\begin{figure}[!t]\centerline{\includegraphics[width=0.27\textwidth]{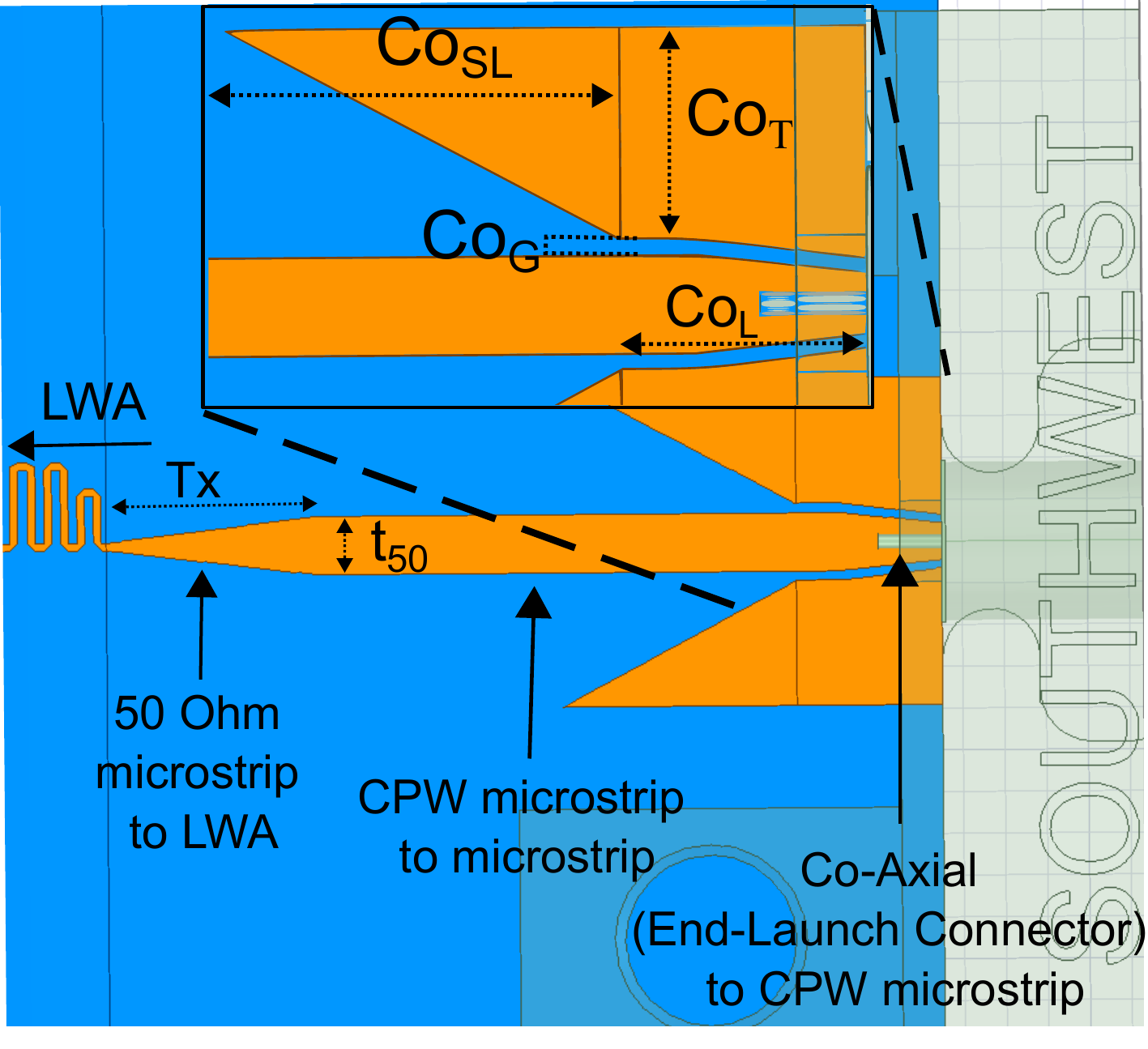}}
\caption{Design of the antenna feed.}
\label{FeedDesign}
\end{figure}

\begin{figure*}[!t]\centerline{\includegraphics[width=0.96\textwidth]{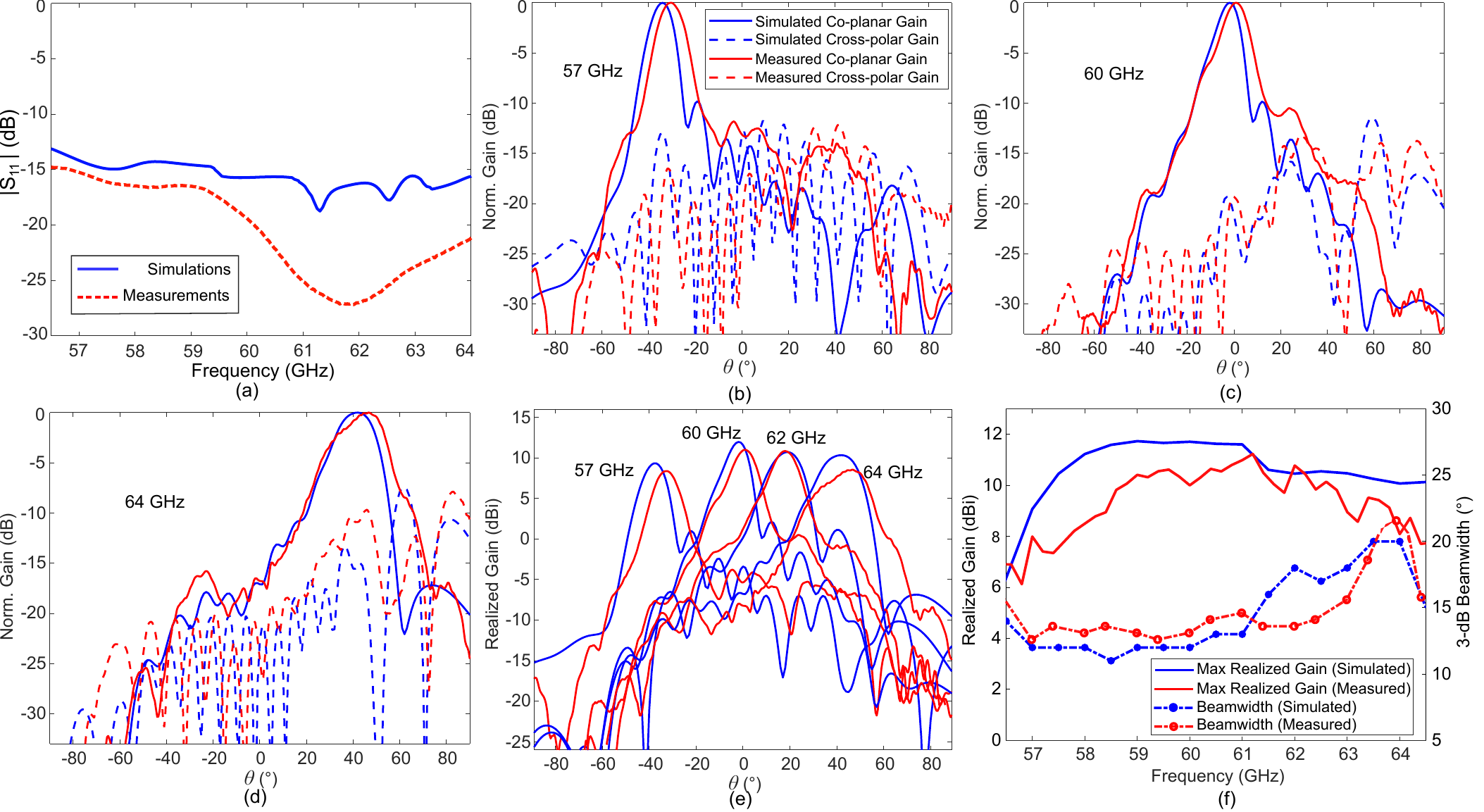}}
\caption{\textcolor{black}{(a)~Simulated and Measured S parameters for the planar condition.} Co-polar realized gain in the H-plane for frequencies $f=$ (b)~\SI{57}{\GHz} (c)~\SI{60}{\GHz} (d)~\SI{64}{\GHz} \textcolor{black}{(e)~Realized Gain at different frequencies} \textcolor{black}{(f)~Simulated and measured realized gain and beamwidth for planar conditions.}}
\label{RadPattern}
\end{figure*}

\begin{figure*}[!t]\centerline{\includegraphics[width=0.96\textwidth]{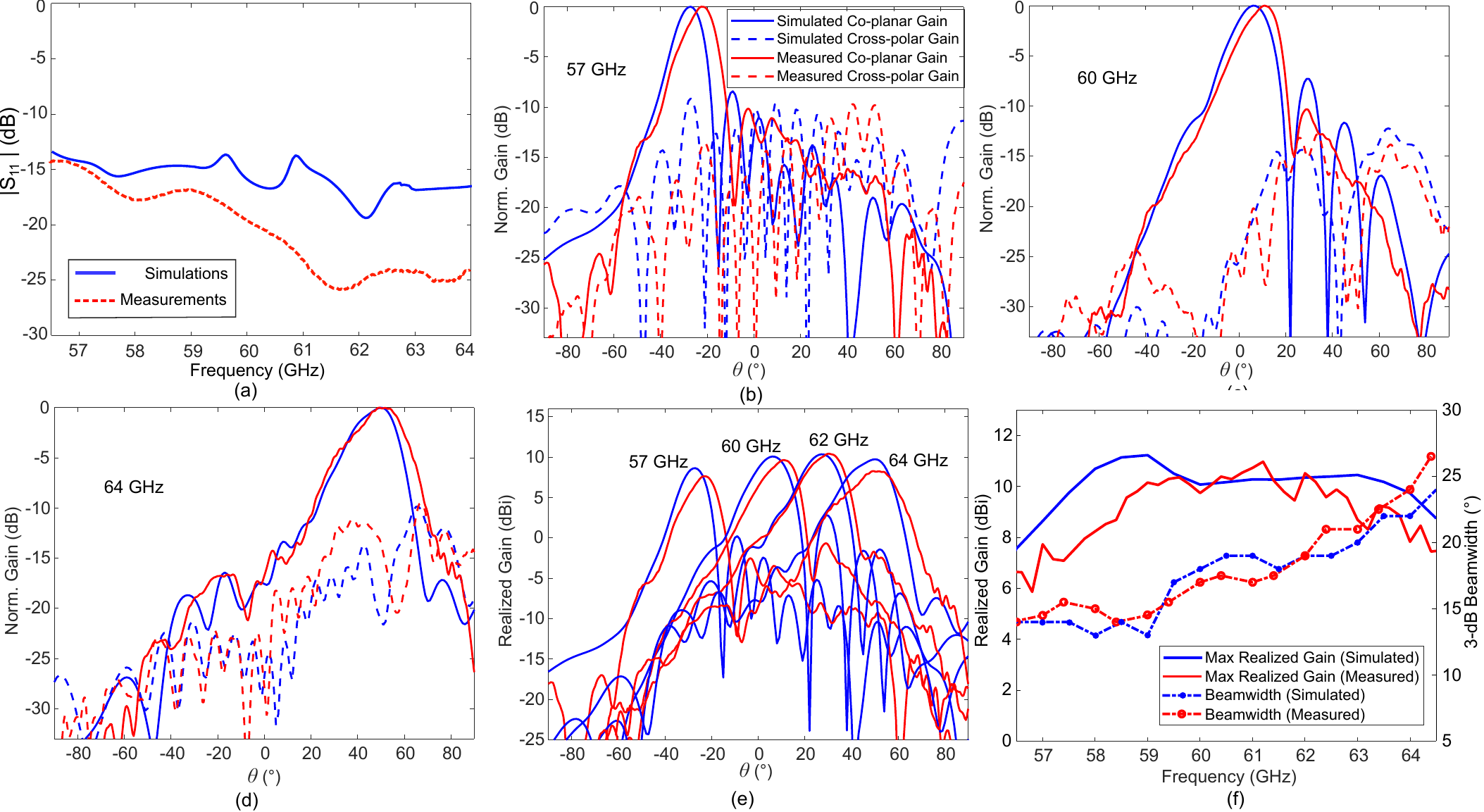}}
\caption{\textcolor{black}{(a)~Simulated and measured S parameters for the bent convex condition with radius of curvature ${R}=$ \SI{80}{\milli\meter}.} Co-polar realized gain in the H-plane for frequencies $f=$ (b)~\SI{57}{\GHz} (c)~\SI{60}{\GHz} (d)~\SI{64}{\GHz} \textcolor{black}{(e)~Realized Gain at different frequencies} \textcolor{black}{(f)~Simulated and measured realized gain and beamwidth for convex conditions.}}
\label{RadPatternCurved}
\end{figure*}

\begin{figure*}[!t]
    \centering
    \hfill
    \begin{subfigure}[b]{0.312\textwidth}
        \centering
        \includegraphics[width=\textwidth]{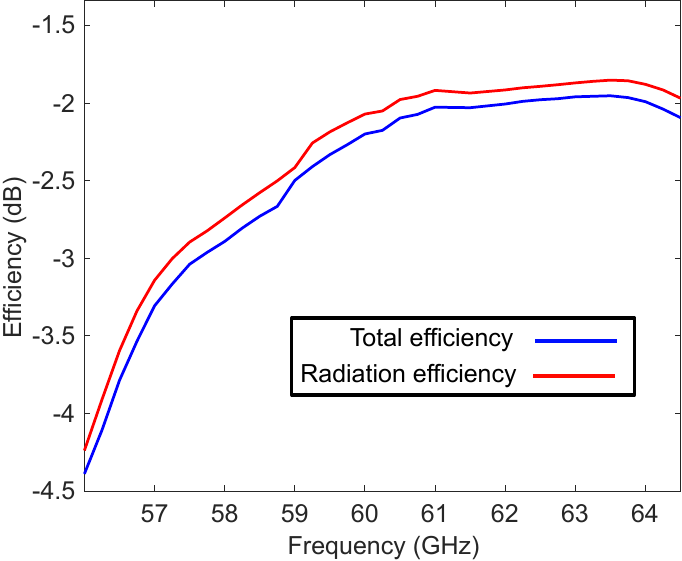}
        \caption{}
        \label{fig:Efficicency}
    \end{subfigure}
    \hfill
    \begin{subfigure}[b]{0.325\textwidth}
        \centering
        \includegraphics[width=\textwidth]{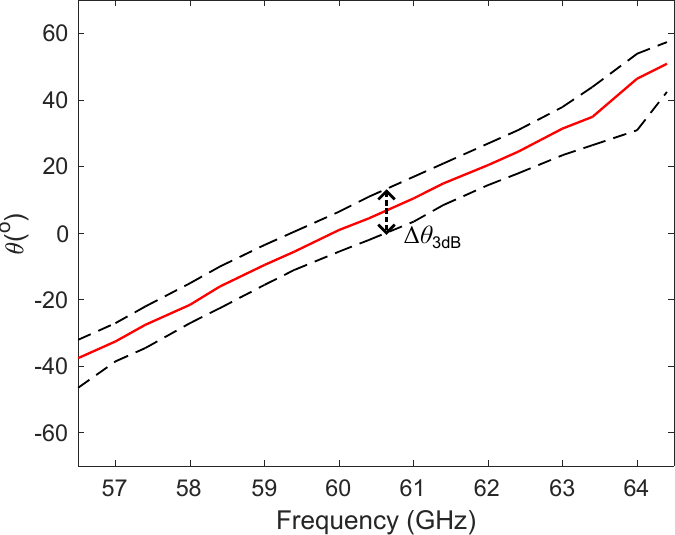}
        \caption{}
        \label{fig:AngularScanningStraight}
    \end{subfigure}
    \hfill
    \begin{subfigure}[b]{0.325\textwidth}
        \centering
        \includegraphics[width=\textwidth]{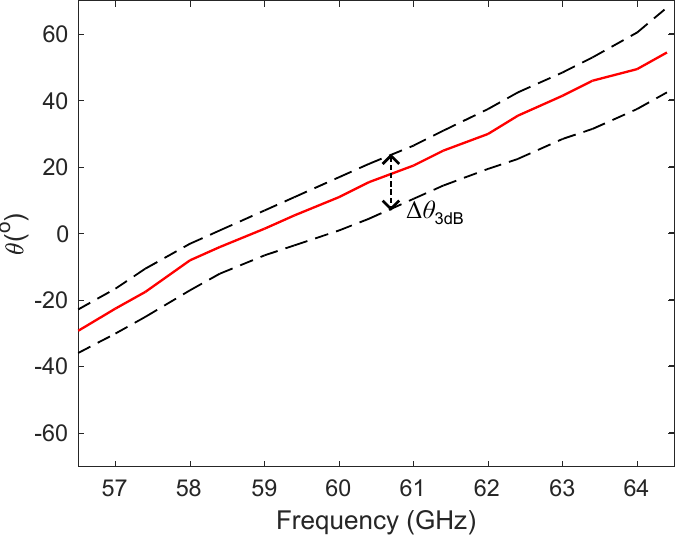}
        \caption{}
        \label{fig:AngularScanningConvex}
    \end{subfigure}
    \caption{(a)~ Radiation and total efficiency of the proposed antenna. \textcolor{black}{Variation of the realized gain as a function of frequency for (b)~planar, and (c)~convex bent condition obtained from measurements. The dotted line indicates the 3-\SI{}{\dB} beamwidth. $\theta_\mathrm{scan}$ shows the scanning domain of the antenna array.} }
    \label{fig:AngularScanning}
\end{figure*}

The designed \gls{lwa} is fed \textit{via} a \SI{1.85}{\milli\meter} end-launch Southwest connector. The feed is designed to avoid the use of vias in the transition from connector (coaxial) to microstrip. Hence a conductor-backed co-planar waveguide transition is used as shown in Fig.~\ref{FeedDesign}. This is done to avoid the spurious radiation occurring during the transition from co-axial feed to the microstrip. The feed is designed to provide optimal performance in the antenna operating range (\SIrange{57}{64}{\GHz}). The dimensions are listed in the Table~\ref{FeedTerminatingPatchDimensions}.

\subsection{Performance of the Designed Prototype}
The designed antenna is prototyped and the \gls{catr} measurements are performed at the CAMILLE facilities of IETR.
A photo of the prototyped \gls{lwa} and the measurement setup are shown in Fig.~\ref{PhotoPrototype} and Fig.~\ref{PhotoPrototypeMeasure}, respectively. The length of the prototyped \gls{lwa} is 5.77$\lambda_\mathrm{d}$, $\lambda_\mathrm{d}$ being the free-space wavelength at $f_\mathrm{d}$).
\subsubsection{In planar conditions}

 the $\vert\mathrm{S}_{11}\vert$ is shown in  Fig.\ref{RadPattern}a indicating that the antenna with the feed remains well matched in the band of operation. As mentioned earlier, the \gls{lwa} is designed to be linearly polarized. This can be observed in the radiation pattern in Fig.\ref{RadPattern}(b--e). The \gls{lwa} scans from \SI{-40}{\degree} to \SI{43}{\degree} in the simulations, while  from \SI{-35}{\degree} to \SI{45}{\degree} experimentally in the range \SIrange[range-units=single, range-phrase=--]{57}{64}{\GHz}. The 3-$\mathrm{dB}$ beamwidth ($\Delta\theta_{3\mathrm{dB}}$) throughout the operational band remains below \SI{20}{\degree}. The maximum side-lobe level remains below \SI{15}{\dB} throughout the operational band. $S_\mathrm{m}$ is \SI[per-mode = symbol]{12.1}{\degree\per\GHz}, while the realized gain stays above \SI{10}{\dB} across the operational range as shown in Fig.~\ref{RadPattern}f.

\subsubsection{In conformal Conditions}
the performance of the antenna in the bent condition is investigated corresponding to the conformal placement on the knee (for $R = $ \SI{80}{\milli\meter}) as shown in Fig.~\ref{FigureOne}~(b). The corresponding $\vert\mathrm{S}_{11}\vert$ is plotted in Fig.~\ref{RadPatternCurved}a indicating that the antenna remains well matched. The normalized realized gain is reported in Fig.~\ref{RadPatternCurved}(b--e). The beam-forming operation remains intact throughout the operational band, while the realized gain decreases up to \SI{1.5}{\dB} when the $R=$ \SI{80}{\milli\meter} as shown in Fig.\ref{RadPatternCurved}f.

The 3-$\mathrm{dB}$ beamwidth varies from \SI{15}{\degree} to \SI{23}{\degree} as depicted in Fig.~\ref{RadPatternCurved}f. Additionally, a shift in the angular scanning domain in bent condition, when compared to the planar condition \cite{martinez-ros_conformal_2014}, can be observed in Fig.~\ref{fig:AngularScanning}(a--b). The change is less than \SI{8}{\degree} across the band of operation.

The radiation and total efficiency of the proposed \gls{lwa} are shown in Fig.~\ref{fig:Efficicency} varying from \SI{-3.3}{\dB} to \SI{-2}{\dB} for the planar and conformal case. \textcolor{black}{The total radiation efficiency of the
antenna is obtained by dividing the measured realized gain
with directivity \cite{sarkar_compact_2020}}.

\subsection{Literature Comparison}
\textcolor{black}{For on-body applications, the antenna must be ergonomic, affordable, and easy to manufacture. The proposed \gls{lwa} consists in a via-free meandering microstrip design with control over scanning rate and gain. The antenna is compact and based on a flexible substrate Rogers 3003.}  
In \cite{steeg_novel_2016, torabi_miniaturized_2023, neophytou_design_2018}, the authors propose a frequency scanning \gls{lwa} based on \gls{siw} technology. However, the $S_\mathrm{m}$ is limited to \SI[per-mode = symbol]{2.15}{\degree\per\GHz}, \SI[per-mode = symbol]{4.45}{\degree\per\GHz}, and \SI[per-mode = symbol]{8.77}{\degree\per\GHz} respectively. \Glspl{lwa} proposed in the works \cite{chang_60-ghz_2016, sarkar_60_2020, 9247092} demonstrate higher or comparable scanning rates than the one proposed in this work, however, these antennas do not offer control over the scanning rate and the realized gain at the same time. A guiding medium created using the meander shaped \gls{siw} is proposed in \cite{shoykhetbrod_design_2012, shoykhetbrod_scanning_2014} to provide additional control over scanning rate, but the angular coverage is limited to \SI{65}{\degree} over the \SIrange[range-units=single, range-phrase=--]{57}{64}{\GHz}. The fast-scanning \gls{lwa} presented in  \cite{jiang_backward--forward_2021, paulotto2014self} operate in the frequency range below \SI{10}{\GHz} and scaling the \gls{lwa} will present manufacturing challenges. Conformal analysis has been performed in \cite{celenk_frequency_2021}, however the $S_\textrm{m}$ is quite low (\SI[per-mode = symbol]{3.23}{\degree\per\GHz}) and the scanning occurs over a large frequency range spanning \SIrange[range-units=single, range-phrase=--]{28}{62}{\GHz}. Additionally, these \Glspl{lwa} utilize \gls{siw} or multi-layer guiding structures, increasing the fabrication complexity. \textcolor{black}{Table~\ref{comparestateoftheart} shows the literature comparison of frequency scanning \glspl{lwa} with similar on-body application.}

\textcolor{black}{The total scanning range of the \gls{lwa} is around $80 \degree$ for both planar ($-35 \degree$ to $45 \degree$) and conformal case  for the planar case ($-25 \degree$ to $55 \degree$) as shown in Fig.~\ref{fig:AngularScanning}(b--c) respectively. The scanning is comparable to commercially available hardware \cite{BGT60LTR11AIP} operating in the \SI{60}{\giga\hertz} band. In addition, the multi-path effects are reduced due to the directional nature of radiaton pattern of the \gls{lwa}.}

\begin{table*}[!t]
\centering
\caption{\textcolor{black}{Comparison of different LWA structures for on-body applications}}
\begin{tabular}{l p{5.5cm} c p{3cm} c}
\toprule
\textbf{Ref.} & \textbf{Structure description} & \textbf{Operating band (GHz)} & \textbf{Angular scanning ($^\circ$)} & \textbf{Gain (dBi)} \\
\midrule
\cite{chen_robust_2024} & \Gls{sspp} based \gls{lwa} on \gls{ebg} ground & 4.7--6 & \SIrange{-15}{85}{\degree} & 11--13 \\
\cite{celenk_frequency_2021} & \gls{siw} transverse slotted \gls{lwa} & 28--62 & \SIrange{-80}{30}{\degree} & 10--25 \\
\cite{kwiatkowski_concept_2017} & Meandering and slotted \gls{siw}-based \gls{lwa} & 70--90 & \SIrange{-25}{30}{\degree} & -- \\
\cite{orth_novel_2019, kwiatkowski_combining_2024} & Same architecture as above & 70--90 & \SIrange{-45}{0}{\degree} & -- \\
\cite{zhang2019high} & 1-D \gls{lwa} based on goubou line architecture loaded with patches & 8.5--10.7 & \SIrange{-70}{70}{\degree} & 8--10.5 \\
{\cite{jiang_backward--forward_2021}} &\textcolor{black}{\gls{lwa} with meandering line structures with inductive open stubs} & 5.9--7.1 & \textcolor{black}{\SIrange{-60}{58}{\degree}} & \textcolor{black}{14--15} \\
\cite{vadher_meandering_2024} & Dual-band \gls{lwa} with novel unit cell based on meandering microstrips & \begin{tabular}{@{}c@{}}19.4--27.5,\\ 11--15.5\end{tabular} & \begin{tabular}{@{}c@{}}\SIrange{-42}{30}{\degree},\\ \SIrange{-15}{60}{\degree}\end{tabular} & \begin{tabular}{@{}c@{}}7--13,\\ 3.1--6\end{tabular} \\
\textbf{This work} & Meandering microstrip \gls{lwa} with control over scanning rate and gain & 57--64 & \begin{tabular}{@{}c@{}}\SIrange{-35}{45}{\degree} (Planar),\\ \SIrange{-25}{55}{\degree} (Conformal)\end{tabular} & 9--10.5 \\
\bottomrule
\end{tabular}
\label{comparestateoftheart}
\end{table*}

\begin{figure}[!t]\centerline{\includegraphics[width=0.25\textwidth]{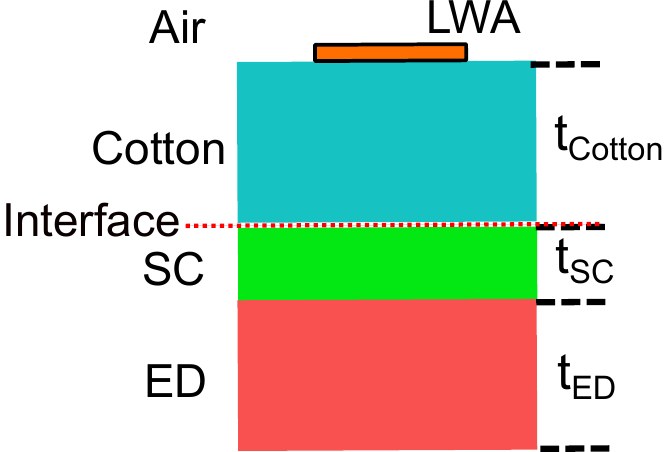}}
\caption{\textcolor{black}{Tissue layers used in the simulation domain with conformal LWA.}}
\label{fig: SimDomain}
\end{figure}

\textcolor{black}{\subsection{On-body Placement of the Antenna}} \label{sec: Onbody}
\textcolor{black}{To test the feasibility of the designed prototype for on-body conditions simulations are performed in the presence of a planar multilayer model of human tissues covered by a layer of textile (cotton). The tissue model and the thickness of each layer is referred from \cite{sacco_impact_2021, ziskin_tissue_2018}.
The simulation domain with the \gls{lwa} and the tissue layer model is shown in Fig.~\ref{fig: SimDomain}. Given the shallow penetration depth at \glspl{mmw}, at \SI{60}{\GHz} only roughly $0.3\%$ of the electromagnetic power reaches the
fat layer underlying the dermis \cite{sacco2021age}. As a consequence, the simulation domain is limited to the dermis layer\cite{noauthor_guidelines_2020, 10568574}. The choice of thickness of each of the layer is motivated by the studies~\cite{sacco_impact_2021, soares_wireless_2023, nikolayev_electromagnetic_2018}.}

\textcolor{black}{Table~\ref{TableTissueProperties} shows the numerical values of permittivity in the operating frequency range and the thickness of each layer. The values of permittivity for tissues are referred from \cite{gabriel_compilation_1996}. Two aspects of on-body placement of the \gls{lwa} have been studied in this section.}

\begin{table}[!t]
    \centering
    \caption{\textcolor{black}{Dielectric properties and thickness of various layers at different frequencies.}}
    \captionsetup{skip=5pt}  
    \begin{tabular}{|c|c|c|c|c|}
        \hline
        \textbf{Layer} & \textbf{Thick.} & \multicolumn{3}{c|}{$\epsilon^*$} \\
        \cline{3-5}
        & \textbf{(mm)} & \SI{57}{\GHz} & \SI{60}{\GHz} & \SI{64}{\GHz} \\
        \hline
        SC & 0.015 & $3.17 - j0.52$ & $3.15 - j0.50$ & $3.12 - j0.48$ \\
        E + D & 1.45 & $8.05 - j11.0$ & $7.98 - j10.9$ & $7.92 - j10.8$ \\
        air & -- & 1 & 1 & 1 \\
        cotton & 0.78 & $2 - j0.04$ & $2 - j0.04$ & $2 - j0.04$ \\
        \hline
    \end{tabular}
    
    \smallskip 
    
    \makebox[\linewidth][r]{\footnotesize {* $\mathrm{E}$ + $\mathrm{D}$ refers to the viable dermis and dermis.}} 
    \label{TableTissueProperties}
\end{table}

\begin{table}[!t]
    \centering
    \caption{\textcolor{black}{Power density at different frequencies with units}}
    \begin{tabular}{|c|c|c|c|}
        \hline
        & \SI{57}{\GHz} & \SI{60}{\GHz} & \SI{64}{\GHz} \\
        \hline
        $\mathrm{APD_{peak, avg, 1 cm^2}}$ (W/m\(^2\)) & 1.9397 & 2.6183 & 5.3668 \\
        $\mathrm{APD_{peak, avg, 4 cm^2}}$ (W/m\(^2\)) & 1.1515 & 1.4386 & 2.7481 \\
        \hline
    \end{tabular}
    \label{APDTAPTable}
\end{table}

\textcolor{black}{\textit{1.~Electromagnetic exposure due to the \gls{lwa} on the body of the user---}}\textcolor{black}{According to the revised guidelines in from IEEE \cite{noauthor_ieee_2019} and ICINIRP \cite{noauthor_guidelines_2020} in 2019 and 2020 respectively, above \SI{6}{\giga\hertz}, the \gls{apd} is used as the \gls{br} and can be computed according to} 

\begin{equation}
\mathrm{APD} = \iint_A \operatorname{Re}[\mathbf{S}] \cdot \, d\mathbf{s} / A = \iint_A \operatorname{Re}[\mathbf{E} \times \mathbf{H}^*] \cdot \, d\mathbf{s} / A\
,
\label{eqS_ab}
\end{equation}
\textcolor{black}{where $\mathbf{E}$ and $\mathbf{H}$ indicate root mean square (rms) values of electric field and magnetic field at the interface respectively, \( \operatorname{Re}[X] \) and \( X^* \) are the real part and the complex conjugate of a complex value \( X \), respectively, and \( d\mathbf{s} \) is the integral variable vector with its direction normal to the integral area \( A \) on the body surface. The \gls{apd} is calculated at the interface between the cotton layer and the \gls{sc} as shown in the Fig.~\ref{fig: SimDomain}. The \gls{apd} limits in the guidelines are given as \SI{20}{\watt\per\square\meter} for an averaging area of \SI{4}{\centi\meter\squared} above \SI{6}{\GHz}. Additional limits of \SI{40}{\watt\per\square\meter} needs to be simultaneously satisfied for the \gls{apd} averaged on \SI{1}{\centi\meter\squared} above \SI{30}{\GHz}. }

\textcolor{black}{In the simulation, a textile thickness $\mathrm{t_{cotton}}=$~\SI{0.78}{\milli\meter} is selected since it corresponds to the maximum \gls{apd} as reported in  \cite{sacco_impact_2021}. To reduce the effects of exposure for \gls{srd} devices FCC \cite{federal2023code} and ETSI \cite{ETSISRD2017} indicate \gls{eirp} limit of \SI{20}{dBm}. This imposes the restriction of maximum input power of \SI{10}{dBm} (ignoring cable losses) since the realized gain of the antenna is around \SI{10}{dB}. The $\mathrm{APD_{peak, avg, 4 cm^2}}$ and $\mathrm{APD_{peak, avg, 1 cm^2}}$ are mentioned in the Table~\ref{APDTAPTable}, which are below the \gls{br}. The ground plane of the \gls{lwa} limits the radiation occurring towards the body and hence minimizes the exposure.}

\begin{figure}[!t]\centerline{\includegraphics[width=0.46\textwidth]{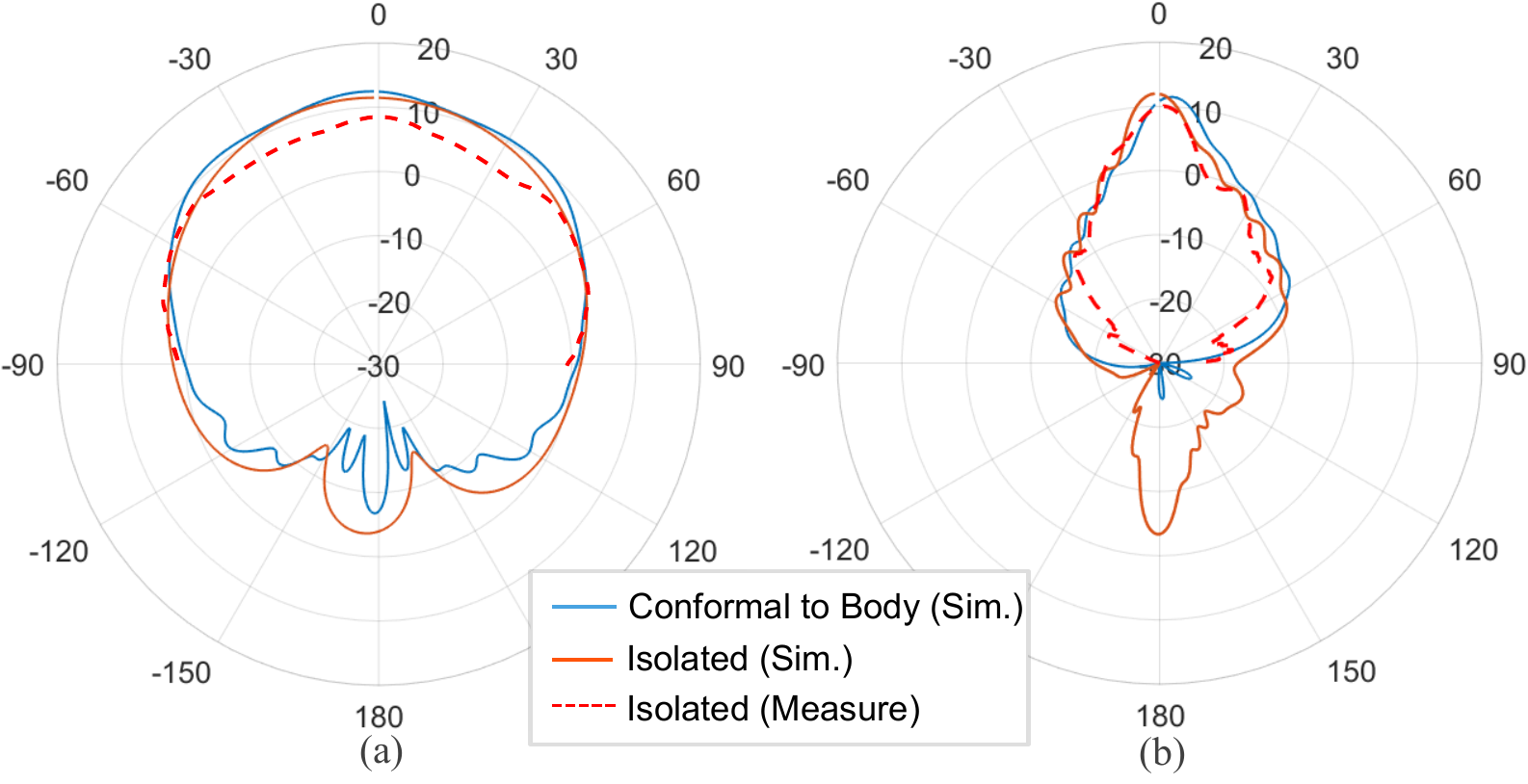}}
\caption{\textcolor{black}{Realized gain pattern in the H-plane and the E-plane respectively at the frequency of broadside radiation $f_\mathrm{d}=$ \SI{60}{\giga\hertz}}.} 
\label{fig: RadPattern_Phi0}
\end{figure}

\textcolor{black}{\textit{2.~Effect on the far-field radiation pattern of the antenna---}}
\textcolor{black}{Fig.~\ref{fig: RadPattern_Phi0}(a--b)  shows the radiation pattern in the H-plane and the E-plane respectively along with the measurements at $f_\mathrm{d}~=~$\SI{60}{\giga\hertz}. Owing to the presence of ground plane, it can be observed that the effect of on-body placement of the antenna on the far-field radiation pattern is minimal.}

\begin{figure*}[!t]\centerline{\includegraphics[width=0.865\textwidth]{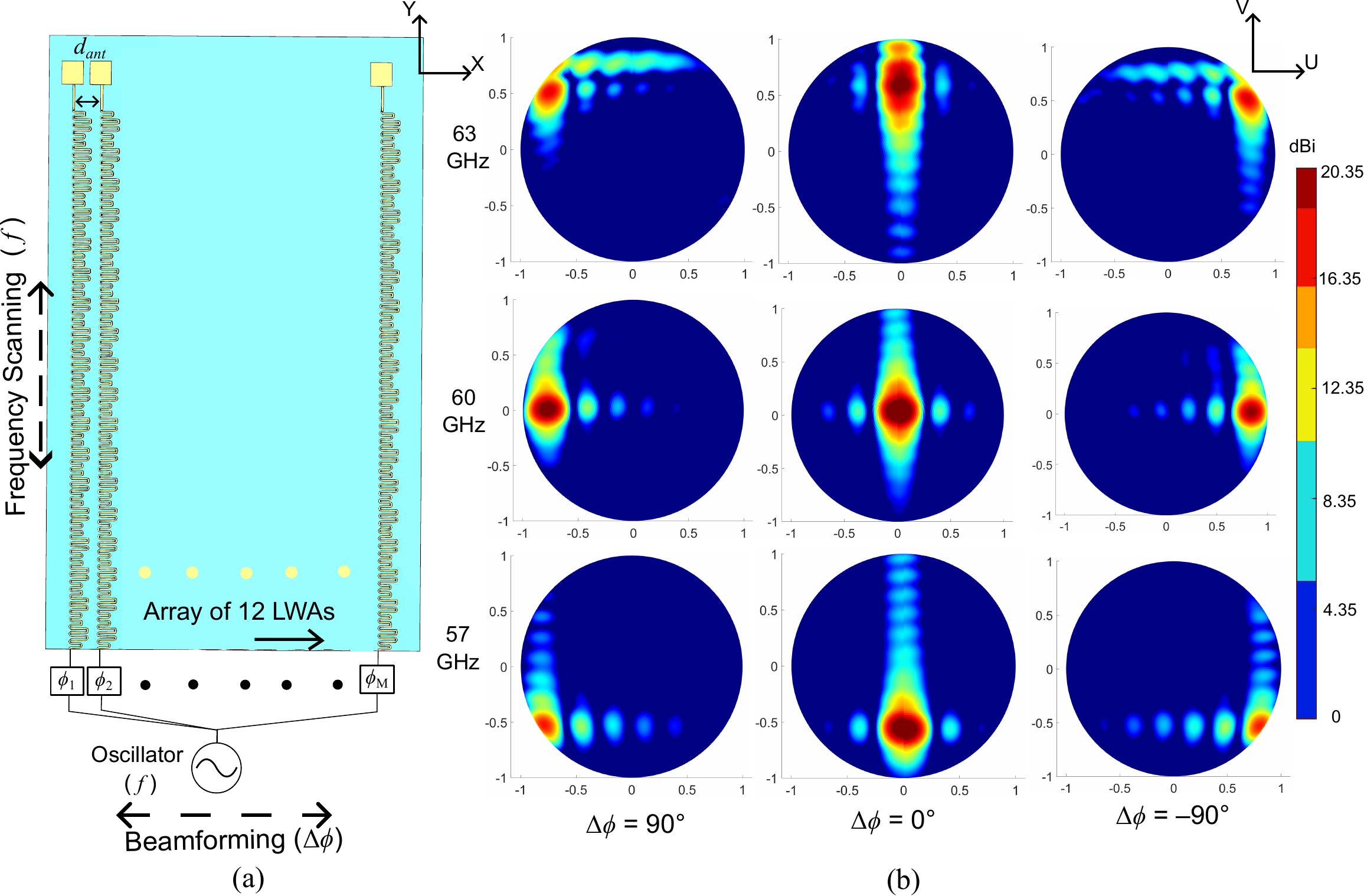}}
\caption{\textcolor{black}{(a)~Array of M$=$12 LWAs for pencil beam scanning with phase shifters. (b)~U--V plane of the normalized realized gain showing the scanning in 2-D.}}
\label{fig:UVSpace}
\end{figure*}

\begin{figure*}[!t]
    \centering
    \begin{subfigure}[b]{0.385\textwidth}
        \centering
            \includegraphics[width=\textwidth]{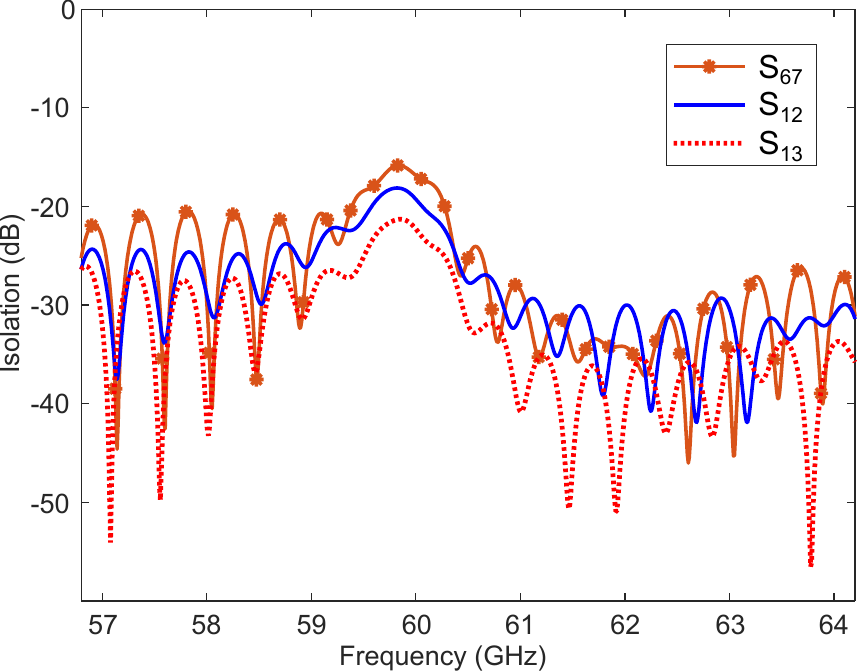}
        \caption{}
        \label{fig: Isolation}
    \end{subfigure}
    \begin{subfigure}[b]{0.375\textwidth}
        \centering
        \includegraphics[width=\textwidth]{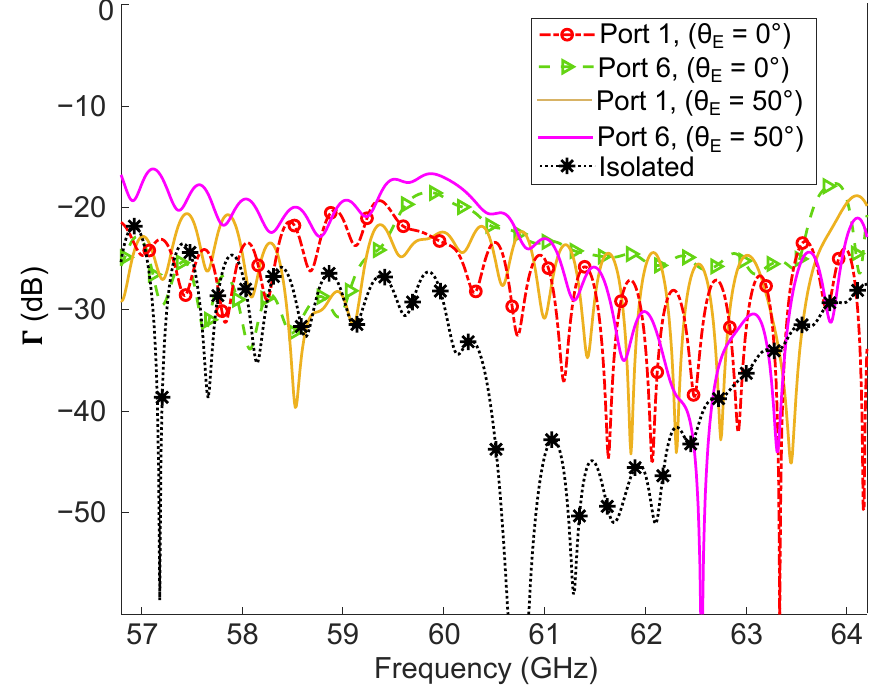}
        \caption{}
        \label{fig: ActiveSPara}
    \end{subfigure}
    \hfill
    \caption{\textcolor{black}{(a)~Isolation between the neighbouring antenna elements. (b)~Active reflection coefficient of the array with reflection coefficient for different ports and scanning angles in E-plane.}}
    \label{}
\end{figure*}

\vspace{0.1in}

\section{Application Example: Array of \Glspl{lwa} for~2-D~Scanning} \label{sec: 2-D scanning}
The proposed \gls{lwa}, as shown in Fig.~\ref{FigureOne}b, radiates a beam that allows scanning in the X--Z plane. If 2-D scanning with a pencil beam is required (Fig.~\ref{FigureOne}a), an array of \glspl{lwa} using beam-forming phased excitation can be employed. 

Fig.~\ref{fig:UVSpace}a shows the design of a 12 \glspl{lwa} array that can be implemented on a single \gls{pcb}. This arrangement enables a pencil-beam scanning in the X--Z and Y--Z planes as shown in  Fig.~\ref{fig:UVSpace}b. The \glspl{lwa} are placed at a distance $d_\text{{ant}} = $ \SI{1.5}{\milli\meter}. The phase difference between each pair can be varied to obtain beamforming in the X--Y plane (E-plane). \textcolor{black}{In Fig.~\ref{fig:UVSpace}b, each port is excited with a phase difference of $\pi/2$ to obtain scanning angle of $\theta_\mathrm{E}=\pm50^\degree$ with a 3-$\mathrm{dB}$ beamwidth of $21\degree$.}

The realized gain radiation pattern for different frequencies and phase excitations is plotted in Fig.~\ref{fig:UVSpace}b projected in the U--V plane ($\mathrm{U}=\sin\theta\cos\phi$, $\mathrm{V}=\sin\theta\sin\phi$).

\subsection{\textcolor{black}{Scattering Parameters of the Array of 12 \glspl{lwa}}}
\textcolor{black}{Fig.~\ref{fig: Isolation} shows the transmission coefficients $\mathrm{S}_{ij}$  in the simulation indicating the isolation between the ports of \gls{lwa}. The twelve-port transmission coefficient is marked as $\mathrm{S}_{ij}$ with (${i, j} \in [1,12]$, $i \neq j$). The transmission coefficients for only the first two pair of ports and the highest transmission coefficients ports (6 and 7) is shown. Results show that the isolation loss is promising for the application, being lower than \SI{-15}{\dB}.}

\subsection{\textcolor{black}{Active Reflection Coefficients in a 12-Port Antenna System}}

\textcolor{black}{In a multi-antenna system, such as a 12-port antenna array described above, the mutual coupling effects between array elements, cannot be just expressed in the form of scattering parameters. The radiation of one element contributes to the other element and therefore induces the coupling between the array elements \cite{balanis_antenna_2015}. The radiation from an array can be better expressed in the form of active reflection coefficients which take into account mutual coupling effects \cite{1229923, 310010}.}

\textcolor{black}{The active reflection coefficient for the $m^\mathrm{th}$ element can be expressed as \cite{310010}.}

\begin{equation}
\Gamma_m(\theta_\mathrm{E}) = e^{jk_0md_{ant} \sin \theta_\mathrm{E}} \sum_{n=1}^{\mathrm{M}} \mathrm{S}_{mn} e^{-jk_0nd_{ant} \sin \theta_\mathrm{E}}    
\end{equation}

\textcolor{black}{Here $\mathrm{M}$ indicates the total number of ports in the \gls{lwa} array. Note that $\Gamma_m$, depends on the scan angle $\theta_\mathrm{E}$, unless $\mathrm{S}_{mn}=0$, $m\neq n$. Fig~\ref{fig: ActiveSPara} shows the active reflection reflection coefficients for two scanning angles in the E-plane ($\theta_\mathrm{E}$). The reflection coefficients for the first port and the reflection coefficient for the sixth port (highest) are shown. The isolated case is also shown to compare and depict the impact of mutual coupling between the arrays.}

\section{Conclusion}\label{SectionConclusion}
\textcolor{black}{This work presents a single-layer, low-profile V-band \gls{lwa} designed for on-body applications. The proposed meandering microstrip structure enables precise control over the scanning rate and realized gain while maintaining a flexible, via-free architecture. The antenna demonstrates rapid frequency scanning and robust performance across planar and conformal conditions, with a scanning range of $-35\degree$ to $45\degree$ in planar form and $-25\degree$ to $55\degree$ when bent to a radius of \SI{80}{\mm}. The presence of a ground plane minimizes user exposure, ensuring compliance with international safety guidelines.}

\textcolor{black}{A possible configuration for two-dimensional scanning is also explored, incorporating mutual coupling effects to assess its feasibility. Results confirm that the beam-forming capabilities remain intact under conformal conditions, albeit with a minor reduction in peak realized gain ($\sim1.5$~dB). Additionally, the antenna’s compatibility with standard PCB fabrication facilitates practical integration into wearable radar systems. Overall, the proposed LWA design offers a compelling solution for wearable millimeter-wave radar applications, balancing performance, manufacturability, and ergonomic constraints.}

\section*{Acknowledgment}
The authors would like to express their gratitude to Zvonimir~Šipuš for his advice on the manuscript and for sharing his insights on periodic structures and stop bands, and to Christos Monochristou for his technical insights into leaky-wave antennas and periodic structures.

\bibliographystyle{IEEEtran}
\bibliography{ZoteroLibrary1107}

%

\mycomment{
\begin{IEEEbiography}[{\includegraphics[width=1in,height=1.25in,clip]{BiographyPhoto/BioPhoto.jpg}}]{Pratik Vadher}
received his M.Tech and B.Tech degrees in Electrical Engineering from the Indian Institute of Technology, Kanpur (IIT-K), Uttar Pradesh, India in 2018. After graduation, he worked at Honeywell Technologies, Bangalore as a Senior Embedded Engineer in the Avionics division for 3 years.

Since 2021, he is pursuing his Ph.D. at the Institute of Electronics and Telecommunication of Rennes (IETR) in Rennes, France. He is a member of the Institute of Electrical and Electronics Engineering (IEEE), the Antennas and Propagation Society (AP-S), and the Microwave Theory and Technology Society (MTT-S). His current research interests encompass periodic structures, conformal antennas, bio-electromagnetics, and radars.
\end{IEEEbiography}}

\mycomment{
\begin{IEEEbiographynophoto}{John Doe}
Biography text here.
\end{IEEEbiographynophoto}}


\mycomment{\begin{IEEEbiographynophoto}{Jane Doe}
Biography text here.
\end{IEEEbiographynophoto}}

\end{document}